
%
\def\unredoffs{}
\tolerance=1000\hfuzz=2pt
\catcode`\@=11 
\ifx\hyperdef\UNd@FiNeD\def\hyperdef#1#2#3#4{#4}\def\hyperref#1#2#3#4{#4}\def\href#1#2{#2}\fi
\magnification=1200\unredoffs\baselineskip=16pt plus 2pt minus 1pt
\def\Date#1{\vfill\leftline{#1}\tenpoint\supereject%
\footline={\hss\tenrm\hyperdef\hypernoname{page}\folio\folio\hss}}%

{\count255=\time\divide\count255 by 60 \xdef\hourmin{\number\count255}
 \multiply\count255 by-60\advance\count255 by\time
 \xdef\hourmin{\hourmin:\ifnum\count255<10 0\fi\the\count255}
}
\def\date{\number\day.\number\month.\number\year\ at \hourmin}


\def\nolabels{\def\wrlabeL##1{}\def\eqlabeL##1{}\def\reflabeL##1{}}
\def\writelabels{\def\wrlabeL##1{\leavevmode\vadjust{\rlap{\smash%
{\line{{\escapechar=` \hfill\rlap{\sevenrm\hskip.03in\string##1}}}}}}}%
\def\eqlabeL##1{{\escapechar-1\rlap{\sevenrm\hskip.05in\string##1}}}%
\def\reflabeL##1{\noexpand\llap{\noexpand\sevenrm\string\string\string##1}}}
\nolabels

\global\newcount\secno \global\secno=0
\global\newcount\meqno \global\meqno=1
\def\s@csym{}

\def\newsec#1\par{\global\advance\secno by1%
{\toks0{#1}\message{(\the\secno. \the\toks0)}}%
\global\subsecno=0\eqnres@t\let\s@csym\secsym\xdef\secn@m{\the\secno}\noindent
{\bf\hyperdef\hypernoname{section}{\the\secno}{\the\secno.} #1}%
\writetoca{{\string\hyperref{}{section}{\the\secno}{\bf \the\secno\quad}} {\bf #1}}\par%
\nobreak\medskip\nobreak\noindent\ignorespaces}
\def\eqnres@t{\xdef\secsym{\the\secno.}\global\meqno=1\bigbreak\bigskip}
\def\sequentialequations{\def\eqnres@t{\bigbreak}}\xdef\secsym{}

\global\newcount\subsecno \global\subsecno=0
\def\subsec#1\par{\global\advance\subsecno by1%
{\toks0{#1}\message{(\s@csym\the\subsecno. \the\toks0)}}%
\global\subsubsecno=0%
\ifnum\lastpenalty>9000\else\bigbreak\fi
\noindent{\it\hyperdef\hypernoname{subsection}{\secn@m.\the\subsecno}%
{\secn@m.\the\subsecno.} #1}\writetoca{\string\hskip1.45cm
{\string\hyperref{}{subsection}{\secn@m.\the\subsecno}{\secn@m.\the\subsecno.}}
{#1}}\par\nobreak\medskip\nobreak\noindent\ignorespaces}

\def\appendix#1#2{\global\meqno=1\global\subsecno=0\xdef\secsym{\hbox{#1.}}%
\bigbreak\bigskip\noindent{\bf Appendix \hyperdef\hypernoname{appendix}{#1}%
{#1.} #2}{\toks0{(#1. #2)}\message{\the\toks0}}%
\xdef\s@csym{#1.}\xdef\secn@m{#1}%
\writetoca{{\string\hyperref{}{appendix}{#1}{\bf {#1}\quad}} {\bf #2}}%
\par\nobreak\medskip\nobreak}

%
\def\checkm@de#1#2{\ifmmode{\def\f@rst##1{##1}\hyperdef\hypernoname{equation}%
{#1}{#2}}\else\hyperref{}{equation}{#1}{#2}\fi}
\def\eqnn#1{\DefWarn#1\xdef #1{(\noexpand\relax\noexpand\checkm@de%
{\s@csym\the\meqno}{\secsym\the\meqno})}%
\wrlabeL#1\writedef{#1\leftbracket#1}\global\advance\meqno by1}
\def\f@rst#1{\c@t#1a\em@ark}\def\c@t#1#2\em@ark{#1}
\def\eqna#1{\DefWarn#1\wrlabeL{#1$\{\}$}%
\xdef #1##1{(\noexpand\relax\noexpand\checkm@de%
{\s@csym\the\meqno\noexpand\f@rst{##1}1}{\hbox{$\secsym\the\meqno##1$}})}
\writedef{#1\numbersign1\leftbracket#1{\numbersign1}}\global\advance\meqno by1}
\def\eqn#1#2{\DefWarn#1%
\xdef #1{(\noexpand\hyperref{}{equation}{\s@csym\the\meqno}%
{\secsym\the\meqno})}$$#2\eqno(\hyperdef\hypernoname{equation}%
{\s@csym\the\meqno}{\secsym\the\meqno})\eqlabeL#1$$%
\writedef{#1\leftbracket#1}\global\advance\meqno by1}
\def\xeqn{\expandafter\xe@n}\def\xe@n(#1){#1}
\def\xeqna#1{\expandafter\xe@n#1}
\def\eqns#1{(\e@ns #1{\hbox{}})}
\def\e@ns#1{\ifx\UNd@FiNeD#1\message{eqnlabel \string#1 is undefined.}%
\xdef#1{(?.?)}\fi{\let\hyperref=\relax\xdef\next{#1}}%
\ifx\next\em@rk\def\next{}\else%
\ifx\next#1\xeqn#1\else\def\n@xt{#1}\ifx\n@xt\next#1\else\xeqna#1\fi
\fi\let\next=\e@ns\fi\next}

\def\DefWarn#1{\ifx\UNd@FiNeD#1\else
\immediate\write16{*** WARNING: the label \string#1 is already defined ***}\fi}
%
\newskip\footskip\footskip14pt plus 1pt minus 1pt 
\def\footnotefont{\ninepoint}\def\f@t#1{\footnotefont #1\@foot}
\def\f@@t{\baselineskip\footskip\bgroup\footnotefont\aftergroup\@foot\let\next}
\setbox\strutbox=\hbox{\vrule height9.5pt depth4.5pt width0pt}
\global\newcount\ftno \global\ftno=0
\def\foot{\global\advance\ftno by1\def\foot@rg{\hyperref{}{footnote}%
{\the\ftno}{\the\ftno}\xdef\foot@rg{\noexpand\hyperdef\noexpand\hypernoname%
{footnote}{\the\ftno}{\the\ftno}}}\footnote{$^{\foot@rg}$}}
%
%
%
\global\newcount\refno \global\refno=1
\newwrite\rfile
\def\ref{[\hyperref{}{reference}{\the\refno}{\the\refno}]\nref}
\def\nref#1{\DefWarn#1%
\xdef#1{[\noexpand\hyperref{}{reference}{\the\refno}{\the\refno}]}%
\writedef{#1\leftbracket#1}%
\ifnum\refno=1\immediate\openout\rfile=\jobname.refs\fi
\chardef\wfile=\rfile\immediate\write\rfile{\noexpand\item{[\noexpand\hyperdef%
\noexpand\hypernoname{reference}{\the\refno}{\the\refno}]\ }%
\reflabeL{#1\hskip.31in}\pctsign}\global\advance\refno by1\findarg}
\def\findarg#1#{\begingroup\obeylines\newlinechar=`\^^M\pass@rg}
{\obeylines\gdef\pass@rg#1{\writ@line\relax #1^^M\hbox{}^^M}%
\gdef\writ@line#1^^M{\expandafter\toks0\expandafter{\striprel@x #1}%
\edef\next{\the\toks0}\ifx\next\em@rk\let\next=\endgroup\else\ifx\next\empty%
\else\immediate\write\wfile{\the\toks0}\fi\let\next=\writ@line\fi\next\relax}}
\def\striprel@x#1{} \def\em@rk{\hbox{}}
\def\lref{\begingroup\obeylines\lr@f}
\def\lr@f#1#2{\DefWarn#1\gdef#1{\let#1=\UNd@FiNeD\ref#1{#2}}\endgroup\unskip}

\def\addref#1{\immediate\write\rfile{\noexpand\item{}#1}} 
\def\listrefs{\vfill\supereject\immediate\closeout\rfile\writestoppt
\baselineskip=\footskip\centerline{{\bf References}}\bigskip{\parindent=20pt%
\frenchspacing\escapechar=` \input \jobname.refs\vfill\eject}\nonfrenchspacing}
\def\startrefs#1{\immediate\openout\rfile=\jobname.refs\refno=#1}
\def\xref{\expandafter\xr@f}\def\xr@f[#1]{#1}
\def\refs#1{\count255=1[\r@fs #1{\hbox{}}]}
\def\r@fs#1{\ifx\UNd@FiNeD#1\message{reflabel \string#1 is undefined.}%
\nref#1{need to supply reference \string#1.}\fi%
\vphantom{\hphantom{#1}}{\let\hyperref=\relax\xdef\next{#1}}%
\ifx\next\em@rk\def\next{}%
\else\ifx\next#1\ifodd\count255\relax\xref#1\count255=0\fi%
\else#1\count255=1\fi\let\next=\r@fs\fi\next}
%

%
\newwrite\ffile\global\newcount\figno \global\figno=1
\def\fig{fig.~\hyperref{}{figure}{\the\figno}{\the\figno}\nfig}
\def\nfig#1{\DefWarn#1%
\xdef#1{fig.~\noexpand\hyperref{}{figure}{\the\figno}{\the\figno}}%
\writedef{#1\leftbracket fig.\noexpand~\xfig#1}%
\ifnum\figno=1\immediate\openout\ffile=\jobname.figs\fi\chardef\wfile=\ffile%
{\let\hyperref=\relax
\immediate\write\ffile{\noexpand\medskip\noexpand\item{Fig.\ %
\noexpand\hyperdef\noexpand\hypernoname{figure}{\the\figno}{\the\figno}. }
\reflabeL{#1\hskip.55in}\pctsign}}\global\advance\figno by1\findarg}
\def\xfig{\expandafter\xf@g}\def\xf@g fig.\penalty\@M\ {}
\def\figs#1{figs.~\f@gs #1{\hbox{}}}
\def\f@gs#1{{\let\hyperref=\relax\xdef\next{#1}}\ifx\next\em@rk\def\next{}\else
\ifx\next#1\xfig #1\else#1\fi\let\next=\f@gs\fi\next}
%
\def\figin{\epsfcheck\figin}\def\figins{\epsfcheck\figins}
\def\epsfcheck{\ifx\epsfbox\UnDeFiNeD
\message{(NO epsf.tex, FIGURES WILL BE IGNORED)}
\gdef\figin##1{\vskip2in}\gdef\figins##1{\hskip.5in}
\else\message{(FIGURES WILL BE INCLUDED)}%
\gdef\figin##1{##1}\gdef\figins##1{##1}\fi}
\def\DefWarn#1{}
\def\figinsert{\goodbreak\topinsert}
\def\ifig#1#2#3{\DefWarn#1\xdef#1{fig.~\the\figno}
\writedef{#1\leftbracket fig.\noexpand~\the\figno}%
\figinsert\figin{\centerline{#3}}
\smallskip
\leftskip=20pt \rightskip=20pt
\baselineskip12pt\noindent
{{\bf Fig.~\the\figno}\ \ninepoint #2}
\medskip
\global\advance\figno by1\par\endinsert}
\newwrite\lfile
{\escapechar-1\xdef\pctsign{\string\%}\xdef\leftbracket{\string\{}
\xdef\rightbracket{\string\}}\xdef\numbersign{\string\#}}
\def\writedefs{\immediate\openout\lfile=label.defs \def\writedef##1{%
{\let\hyperref=\relax\let\hyperdef=\relax\let\hypernoname=\relax
 \immediate\write\lfile{\string\def\string##1\rightbracket}}}}%
\def\writestop{\def\writestoppt{\immediate\write\lfile{\string\pageno
 \the\pageno\string\startrefs\leftbracket\the\refno\rightbracket
 \string\def\string\secsym\leftbracket\secsym\rightbracket
 \string\secno\the\secno\string\meqno\the\meqno}\immediate\closeout\lfile}}
\def\writestoppt{}\def\writedef#1{}

\def\seclab#1{\DefWarn#1%
\xdef #1{\noexpand\hyperref{}{section}{\the\secno}{\the\secno}}%
\writedef{#1\leftbracket#1}\wrlabeL{#1=#1}}
\def\subseclab#1{\DefWarn#1%
\xdef #1{\noexpand\hyperref{}{subsection}{\the\secno.\the\subsecno}%
{\the\secno.\the\subsecno}}\writedef{#1\leftbracket#1}\wrlabeL{#1=#1}}
\def\applab#1{\DefWarn#1%
\xdef #1{\noexpand\hyperref{}{appendix}{\secn@m}{\secn@m}}%
\writedef{#1\leftbracket#1}\wrlabeL{#1=#1}}
\newwrite\tfile \def\writetoca#1{}
\def\leaderfill{\leaders\hbox to 1em{\hss.\hss}\hfill}
\def\writetoc{\immediate\openout\tfile=\jobname.toc
   \def\writetoca##1{{\edef\next{\write\tfile{\noindent ##1
   \string\leaderfill{
   \string\hyperref{}{page}{\noexpand\number\pageno}%
   {\noexpand\number\pageno}} \par}}\next}}
}
\newread\ch@ckfile
\def\listtoc{\immediate\closeout\tfile\immediate\openin\ch@ckfile=\jobname.toc
\ifeof\ch@ckfile\message{no file \jobname.toc, no table of contents this pass}%
\else\closein\ch@ckfile\centerline{\bf Contents}\nobreak\medskip%
{\baselineskip=16pt\footnotefont\parskip=0pt\catcode`\@=11\input\jobname.toc
\catcode`\@=12\bigbreak\bigskip}\fi}
\catcode`\@=12 
\def\tenpoint{\def\rm{\fam0\tenrm}
\textfont0=\tenrm \scriptfont0=\sevenrm \scriptscriptfont0=\fiverm
\textfont1=\teni  \scriptfont1=\seveni  \scriptscriptfont1=\fivei
\textfont2=\tensy \scriptfont2=\sevensy \scriptscriptfont2=\fivesy
\textfont\itfam=\tenit \def\it{\fam\itfam\tenit}\def\footnotefont{\ninepoint}%
\textfont\bffam=\tenbf \def\bf{\fam\bffam\tenbf}\def\sl{\fam\slfam\tensl}\rm}
\font\ninerm=cmr9 \font\sixrm=cmr6 \font\ninei=cmmi9 \font\sixi=cmmi6
\font\ninesy=cmsy9 \font\sixsy=cmsy6 \font\ninebf=cmbx9
\font\nineit=cmti9 \font\ninesl=cmsl9 \skewchar\ninei='177
\skewchar\sixi='177 \skewchar\ninesy='60 \skewchar\sixsy='60
\def\ninepoint{\def\rm{\fam0\ninerm}
\textfont0=\ninerm \scriptfont0=\sixrm \scriptscriptfont0=\fiverm
\textfont1=\ninei \scriptfont1=\sixi \scriptscriptfont1=\fivei
\textfont2=\ninesy \scriptfont2=\sixsy \scriptscriptfont2=\fivesy
\textfont\itfam=\ninei \def\it{\fam\itfam\nineit}\def\sl{\fam\slfam\ninesl}%
\textfont\bffam=\ninebf \def\bf{\fam\bffam\ninebf}\rm}
%
\hyphenation{anom-aly anom-alies coun-ter-term coun-ter-terms}

\global\newcount\subsubsecno \global\subsubsecno=0
\def\subsubsec#1\par{\global\advance\subsubsecno by1%
{\toks0{#1}\message{(\the\secno\the\subsecno\the\subsubsecno. \the\toks0)}}%
\ifnum\lastpenalty>9000\else\bigbreak\fi
\noindent{\it\hyperdef\hypernoname{subsubsection}{\the\secno.\the\subsecno\the\subsubsecno}%
{\the\secno.\the\subsecno.\the\subsubsecno.} #1}
\par\nobreak\medskip\nobreak\noindent\ignorespaces}

\def\DefWarn#1{}
\def\tikzcaption#1#2{\DefWarn#1\xdef#1{Fig.~\the\figno}
\writedef{#1\leftbracket Fig.\noexpand~\the\figno}%
{
\smallskip
\leftskip=20pt \rightskip=20pt \baselineskip12pt\noindent
{{\bf Fig.~\the\figno}\ \ninepoint #2}
\bigskip
\global\advance\figno by1 \par}}

\def\ntoalpha#1{%
\ifcase#1%
@%
\or A\or B\or C\or D\or E\or F\or G\or H\or I
\fi
}

\global\newcount\appno \global\appno=1
\def\applab#1{\xdef #1{\ntoalpha\appno}\writedef{#1\leftbracket#1}\wrlabeL{#1=#1}
\global\advance\appno by1}

\def\preprint#1 #2\par{\rightline{\vbox{\baselineskip12pt\hbox{#1}\hbox{#2}}}\vskip2cm}
%
\def\title#1\par{\centerline{\bf #1}\nopagenumbers\pageno=0}
\def\author#1\par{\bigskip\bigskip\centerline{#1}}

\newcount\addressno

\def\email#1#2{\unskip$^#1$\footnote{\null}{\kern-\parindent \llap{$^#1$\hskip1pt}email: #2}}

\def\startcenter{%
  \par
  \begingroup
  \leftskip=0pt plus 1fil
  \rightskip=\leftskip
  \parindent=0pt
  \parfillskip=0pt
}
\def\stopcenter{\endgroup}

\def\address{\bigskip%
  \ifnum\the\addressno=0\else\stopcenter\endgroup\fi
  \advance\addressno by 1%
  \begingroup
  \startcenter
  \it
  \obeylines
  \addressAux
}
\def\addressAux#1{#1}

\def\abstract{\stopcenter\endgroup\bigskip\bigskip\noindent}

\def\Dsl{\,\raise.15ex\hbox{/}\mkern-13.5mu D} 
\def\dsl{\raise.15ex\hbox{/}\kern-.57em\partial}
 
\def\boxeqn#1{\vcenter{\vbox{\hrule\hbox{\vrule\kern3pt\vbox{\kern3pt
	\hbox{${\displaystyle #1}$}\kern3pt}\kern3pt\vrule}\hrule}}}


\def\half{{1\over 2}}

\def\bar{\overline}
\def\({\left(}
\def\){\right)}



\def\qed{\hbox{\hskip 3pt
\vbox{\hrule\hbox to 7pt{\vrule height 7pt\hfill\vrule}
\hrule}}\hskip3pt}

\overfullrule=0pt\relax

\frenchspacing

\newread\instream \openin\instream= label.defs
\ifeof\instream \message{No labels in advance yet. Wait till next pass.}
\else \closein\instream \input label.defs
\fi
\writedefs

\def\arXiv:#1].{\hepthStrip#1 \nil}
\def\hepthStrip#1 #2\nil{\href{http://arxiv.org/abs/#1}{arXiv:#1 #2\unskip}].}

\input epsf
\input figflow

\def\centretable#1{ \hbox to \hsize {\hfill\vbox{
                    \offinterlineskip \tabskip=0pt \halign{#1} }\hfill} }

\preprint{}

\vskip -.7in

\title The 11D pure spinor ghost number zero vertex operator

\author Max Guillen\email{{\dagger\ddagger *}}{maxgui@chalmers.se}, Marcelo  dos Santos\email{{*\#}}{mafsantos@ucdavis.edu}, Eggon Viana\email{{*\star\circ}}{eggon.viana@unesp.br}

\address
$^\dagger$ Department of Physics and Astronomy, Uppsala University, 75108 Uppsala, Sweden

\vskip .1in
$\ddagger$ Department of Mathematical Sciences, Chalmers University of Technology and the University of Gothenburg, SE-412 96 Gothenburg, Sweden

\vskip .1in
$^{*}$ ICTP South American Institute for Fundamental Research
Instituto de F\'{i}sica Te\'{o}rica, UNESP-Universidade Estadual Paulista
R. Dr. Bento T. Ferraz 271, Bl. II, S\~{a}o Paulo 01140-070, SP, Brazil

\vskip .1in
${\#}$ Center for Quantum Mathematics and Physics (QMAP)
Department of Physics \& Astronomy, University of California, Davis, CA 95616 USA

\vskip .1in
${\star}$ Instituto Gallego de F\'{i}sica de Altas Energ\'{i}as (IGFAE), Spain

\vskip .1in
${\circ}$ Department of Mathematical Sciences, Durham University, Durham DH1 3LE, UK

\vskip -.1in

\abstract
The 11D pure spinor worldline has been proved to successfully describe the physical states of 11D supergravity in a manifestly super-Poincar\'e covariant fashion. Within this framework, the computation of scattering amplitudes requires the existence of vertex operators carrying different ghost numbers. A recent no-go theorem demonstrated the impossibility of constructing a ghost number zero vertex operator consistent with 11D supergravity in the minimal pure spinor formalism. In this letter, we overcome this obstruction by working in the non-minimal formulation of the 11D pure spinor superparticle. We construct, for the first time, a ghost number zero vertex operator with a remarkably compact structure when expressed in terms of physical operators. We further verify that it satisfies the expected descent relation with the ghost number one vertex operator, and that its commutator with the ghost number three single-particle vertex reproduces the two-particle superfield recently introduced in the literature.

\bigskip
\bigskip
\bigskip
\bigskip
\vskip -.2in
\Date {August 2025}

\newif\iffig
\figfalse



\lref\psparticle{
N.~Berkovits,
``Covariant quantization of the superparticle using pure spinors,''
JHEP {\bf 09}, 016 (2001).
[arXiv:0105050 [hep-th]].
}

\lref\ICTP{
	N.~Berkovits,
  	``ICTP lectures on covariant quantization of the superstring,''
	[arXiv:020 9059 [hep-th]].
}

\lref\tamingbghost{
M.~Guillen,
``Taming the 11D pure spinor b-ghost,''
JHEP {\bf 03}, 135 (2023).
[arXiv:2212. 13653 [hep-th]].
}

\lref\elevendsuperspaceexpansion{
M.~Ben-Shahar and M.~Guillen,
``Superspace expansion of the 11D linearized superfields in the pure spinor formalism, and the covariant vertex operator,''
JHEP {\bf 09}, 018 (2023).
[arXiv:2305.19898 [hep-th]].
}

\lref\pselevenparticle{
M.~Guillen,
``Equivalence of the 11D pure spinor and Brink-Schwarz-like superparticle cohomologies,''
Phys. Rev. D {\bf 97}, no.6, 066002 (2018).
[arXiv:1705.06316 [hep-th]].
}

\lref\superpoincarequantization{
N.~Berkovits,
``Super Poincare covariant quantization of the superstring,''
JHEP {\bf 04}, 018 (2000).
[arXiv:hep-th/0001035 [hep-th]].
}

\lref\berkovitsguillenbghost{
N.~Berkovits and M.~Guillen,
``Simplified $D = 11$ pure spinor $b$ ghost,''
JHEP {\bf 07}, 115 (2017).
[arXiv:1703.05116 [hep-th]].
}

\lref\guillenchiral{
M.~Guillen,
``Green-Schwarz and pure spinor formulations of chiral strings,''
JHEP {\bf 12}, 029 (2021).
[arXiv:2108.11724 [hep-th]].
}

\lref\pssupermembrane{
N.~Berkovits,
``Towards covariant quantization of the supermembrane,''
JHEP {\bf 09}, 051 (2002).
[arXiv:0201151 [hep-th]].
}

\lref\neight{
M.~Cederwall,
``N=8 superfield formulation of the Bagger-Lambert-Gustavsson model,''
JHEP {\bf 09}, 116 (2008).
[arXiv:0808.3242 [hep-th]].
}

\lref\nsix{
M.~Cederwall,
``Superfield actions for N=8 and N=6 conformal theories in three dimensions,''
JHEP {\bf 10}, 070 (2008).
[arXiv:0809.0318 [hep-th]].
}

\lref\nfour{
M.~Cederwall,
``An off-shell superspace reformulation of D=4, N=4 super-Yang-Mills theory,''
Fortsch. Phys. {\bf 66}, no.1, 1700082 (2018).
[arXiv:1707.00554 [hep-th]].
}

\lref\pssugra{
M.~Cederwall,
``D=11 supergravity with manifest supersymmetry,''
Mod. Phys. Lett. A {\bf 25}, 3201-3212 (2010).
[arXiv:1001.0112 [hep-th]].
}

\lref\psborninfeld{
M.~Cederwall and A.~Karlsson,
``Pure spinor superfields and Born-Infeld theory,''
JHEP {\bf 11}, 134 (2011).
[arXiv:1109.0809 [hep-th]].
}

\lref\mafraone{
N.~Berkovits and C.~R.~Mafra,
``Some Superstring Amplitude Computations with the Non-Minimal Pure Spinor Formalism,''
JHEP {\bf 11}, 079 (2006).
[arXiv:hep-th/0607187 [hep-th]].
}

\lref\mafratwo{
H.~Gomez and C.~R.~Mafra,
``The Overall Coefficient of the Two-loop Superstring Amplitude Using Pure Spinors,''
JHEP {\bf 05}, 017 (2010).
[arXiv:1003.0678 [hep-th]].
}

\lref\mafrathree{
H.~Gomez and C.~R.~Mafra,
``The closed-string 3-loop amplitude and S-duality,''
JHEP {\bf 10}, 217 (2013).
[arXiv:1308.6567 [hep-th]].
}

\lref\rnspsone{
N.~Berkovits,
``Covariant Map Between Ramond-Neveu-Schwarz and Pure Spinor Formalisms for the Superstring,''
JHEP {\bf 04}, 024 (2014).
[arXiv:1312.0845 [hep-th]].
}

\lref\rnspstwo{
N.~Berkovits,
``Manifest spacetime supersymmetry and the superstring,''
JHEP {\bf 10}, 162 (2021).
[arXiv:2106.04448 [hep-th]].
}

\lref\maxmaor{
M.~Ben-Shahar and M.~Guillen,
``10D super-Yang-Mills scattering amplitudes from its pure spinor action,''
JHEP {\bf 12}, 014 (2021).
[arXiv:2108.11708 [hep-th]].
}

\lref\stieberger{
C.~R.~Mafra, O.~Schlotterer and S.~Stieberger,
``Complete N-Point Superstring Disk Amplitude I. Pure Spinor Computation,''
Nucl. Phys. B {\bf 873}, 419-460 (2013).
[arXiv:1106.2645 [hep-th]].
}

\lref\mafraoli{
C.~R.~Mafra and O.~Schlotterer,
``Multiparticle SYM equations of motion and pure spinor BRST blocks,''
JHEP {\bf 07}, 153 (2014).
[arXiv:1404.4986 [hep-th]].
}

\lref\dynamical{
N.~Berkovits,
``Dynamical twisting and the b ghost in the pure spinor formalism,''
JHEP {\bf 06}, 091 (2013).
[arXiv:1305.0693 [hep-th]].
}

\lref\xiyin{
C.~M.~Chang, Y.~H.~Lin, Y.~Wang and X.~Yin,
``Deformations with Maximal Supersymmetries Part 2: Off-shell Formulation,''
JHEP {\bf 04}, 171 (2016).
[arXiv:1403.0709 [hep-th]].
}

\lref\chiralmax{
M.~Guillen,
``Green-Schwarz and pure spinor formulations of chiral strings,''
JHEP {\bf 12}, 029 (2021).
[arXiv:2108.11724 [hep-th]].
}

\lref\bcjone{
Z.~Bern, J.~J.~M.~Carrasco and H.~Johansson,
``New Relations for Gauge-Theory Amplitudes,''
Phys. Rev. D {\bf 78}, 085011 (2008).
[arXiv:0805.3993 [hep-ph]].
}

\lref\bcjtwo{
Z.~Bern, J.~J.~M.~Carrasco and H.~Johansson,
``Perturbative Quantum Gravity as a Double Copy of Gauge Theory,''
Phys. Rev. Lett. {\bf 105}, 061602 (2010).
[arXiv:1004.0476 [hep-th]].
}

\lref\bcjthree{
Z.~Bern, J.~J.~Carrasco, M.~Chiodaroli, H.~Johansson and R.~Roiban,
``The Duality Between Color and Kinematics and its Applications,''
[arXiv:1909.01358 [hep-th]].
}

\lref\maximalloopcederwall{
M.~Cederwall and A.~Karlsson,
``Loop amplitudes in maximal supergravity with manifest supersymmetry,''
JHEP {\bf 03}, 114 (2013).
[arXiv:1212.5175 [hep-th]].
}

\lref\maxnotesworldline{
M.~Guillen,
``Notes on the 11D pure spinor wordline vertex operators,''
JHEP {\bf 08}, 122 (2020).
[arXiv:2006.06022 [hep-th]].
}

\lref\OdaTonin{
I.~Oda and M.~Tonin,
``On the Berkovits covariant quantization of GS superstring,''
Phys. Lett. B {\bf 520}, 398-404 (2001).
[arXiv:hep-th/0109051 [hep-th]].
}

\lref\perturbiner{
A.~A.~Rosly and K.~G.~Selivanov,
``On amplitudes in selfdual sector of Yang-Mills theory,''
Phys. Lett. B {\bf 399}, 135-140 (1997).
[arXiv:hep-th/9611101 [hep-th]].
}
\lref\NMPS{
	N.~Berkovits,
	``Pure spinor formalism as an N=2 topological string,''
	JHEP {\bf 10}, 089 (2005).
	[arXiv:0509120 [hep-th]].
}

\lref\elevendsimplifiedb{
N.~Berkovits and M.~Guillen,
``Simplified $D = 11$ pure spinor $b$ ghost,''
JHEP {\bf 07}, 115 (2017).
[arXiv:1703.05116 [hep-th]].
}
\lref\brinkschwarz{
L.~Brink and J.~H.~Schwarz,
``Quantum Superspace,''
Phys. Lett. B {\bf 100}, 310-312 (1981).
}

\lref\brinkhowe{
L.~Brink and P.~S.~Howe,
``Eleven-Dimensional Supergravity on the Mass-Shell in Superspace,''
Phys. Lett. B {\bf 91}, 384-386 (1980).
}

\lref\quartet{
T.~Kugo and I.~Ojima,
``Local Covariant Operator Formalism of Nonabelian Gauge Theories and Quark Confinement Problem,''
Prog. Theor. Phys. Suppl. {\bf 66}, 1-130 (1979).
}

\lref\cederwallequations{
N.~Berkovits and M.~Guillen,
``Equations of motion from Cederwall's pure spinor superspace actions,''
JHEP {\bf 08}, 033 (2018).
[arXiv:1804.06979 [hep-th]].
}

\lref\pssreview{
M.~Cederwall,
``Pure spinor superfields -- an overview,''
Springer Proc. Phys. {\bf 153}, 61-93 (2014).
[arXiv:1307.1762 [hep-th]].
}

\lref\tendsupertwistors{
N.~Berkovits,
``Ten-Dimensional Super-Twistors and Super-Yang-Mills,''
JHEP {\bf 04}, 067 (2010).
[arXiv:0910.1684 [hep-th]].
}

\lref\maxdiegoone{
D.~Garc\'\i{}a Sep\'ulveda and M.~Guillen,
``A pure spinor twistor description of the $D = 10$ superparticle,''
JHEP {\bf 08}, 130 (2020).
[arXiv:2006.06023 [hep-th]].
}

\lref\maxdiegotwo{
D.~G.~Sep\'ulveda and M.~Guillen,
``A Pure Spinor Twistor Description of Ambitwistor Strings,''
[arXiv:2006.06025 [hep-th]].
}

\lref\nmmax{
N.~Berkovits, M.~Guillen and L.~Mason,
``Supertwistor description of ambitwistor strings,''
JHEP {\bf 01}, 020 (2020).
[arXiv:1908.06899 [hep-th]].
}

\lref\maxmasoncasaliberkovits{
N.~Berkovits, E.~Casali, M.~Guillen and L.~Mason,
``Notes on the $D=11$ pure spinor superparticle,''
JHEP {\bf 08}, 178 (2019).
[arXiv:1905.03737 [hep-th]].
}

\lref\maxthesis{
M.~Guillen,
``Pure spinors and $D=11$ supergravity,''
[arXiv:2006.06014 [hep-th]].
}

\lref\measureeleven{
M. Cederwall,
``Towards a manifestly supersymmetric action for 11-dimensional supergravity,''
JHEP 01 (2010) 117, arXiv:0912.1814 [hep-th].
}

\lref\Nahm{
W. Nahm,
``Supersymmetries And Their Representations,''
Nucl. Phys. B135 (1978) 149.
}

\lref\Scherk{
E. Cremmer, B. Julia, and J. Scherk,
``Supergravity Theory In 11 Dimensions,''
Phys. Lett. 76B (1978) 409.
}

\lref\WittenM{
E. Witten,
``String theory dynamics in various dimensions,''
Nucl. Phys. B 443 (1995) 85–126, arXiv:hep-th/9503124.
}

\lref\TownsendM{
C. Hull and P. Townsend,
``Unity of superstring dualities,''
Nucl. Phys. B 438 (1995) 109–137, arXiv:hep-th/9410167.
}

\lref\Duff{
M. Duff, P. S. Howe, T. Inami, and K. Stelle,
``Superstrings in D=10 from Supermembranes in D=11,''
Phys. Lett. B 191 (1987) 70.
}

\lref\Bergshoeff{
E. Bergshoeff, E. Sezgin, and P. Townsend,
``Supermembranes and Eleven-Dimensional Supergravity,''
Phys. Lett. B 189 (1987) 75–78.
}

\lref\tendimensions{
M. Guillen, M. dos Santos, and E. Viana.
``The pure spinor superparticle and 10D super-Yang-Mills amplitudes,'' JHEP {\bf 12}, 044 (2025). [arXiv:2508.19601 [hep-th]].
}

\lref\elevenpart{
M. B. Green, M. Gutperle, and H. H. Kwon,
``Light cone quantum mechanics of the eleven-dimensional superparticle,''
JHEP {\bf 08}, 012 (1999), [arXiv:9907155 [hep-th]].
}

\lref\amplitudeeleven{
M.~Guillen, M.~dos~Santos, E.~Viana
``Tree-level 11D supergravity amplitudes from the pure spinor worldline,'' [arXiv:2508.19744 [hep-th]].

}

\lref\Grassi{
L. Anguelova, P. A. Grassi, and P. Vanhove,
``Covariant one-loop amplitudes in D=11,''
Nucl. Phys. B 702 (2004) 269–306. [arXiv:0408171 [hep-th]].
}

\lref\Karlsson{
A. Karlsson,
``Ultraviolet divergences in maximal supergravity from a pure spinor point of view,''
JHEP {\bf 04}, (2015) 165. [arXiv:1412.5983 [hep-th]].
}

\lref\Berkovitstwoloops{
N. Berkovits,
``Super-Poincare covariant two-loop superstring amplitudes,''
JHEP {\bf 01}, 005 (2006). [arXiv:0503197 [hep-th]].
}

\lref\BerkovitsMafratwoloops{
N. Berkovits and C. R. Mafra,
``Equivalence of two-loop superstring amplitudes in the
pure spinor and RNS formalisms,'' Phys. Rev. Lett. 96, 011602 (2006). [arXiv:0509234 [hep-th]].
}

\lref\GS{
 M.B. Green and J.H. Schwarz,
 ``Covariant Description of Superstrings,''
 Phys. Lett. B136 (1984) 367.
}

\lref\GSsusy{
M. B. Green and J. H. Schwarz,
``Supersymmetrical String Theories,''
Phys. Lett. B 109 (1982) 444.
}

\lref\GSYMsugra{
L. Brink, M. B. Green and J. H. Schwarz,
``N=4 Yang-Mills and N=8 Supergravity as Limits of String Theories,''
Nucl. Phys. B 198, 474-492 (1982).
}

\lref\tenDsYM{
L. Brink, J. H. Schwarz, and J. Scherk,
``Supersymmetric Yang-Mills Theories,''
Nucl. Phys. B 121 (1977) 77–92.
}

\lref\tenLCG{
L. Brink, O. Lindgren, B. Nilsson,
``N=4 Yang–Mills Theory on the Light Cone,''
Nucl. Phys. B212 (1983) 401.
}

\lref\supertwistorambitwistor{
N.~Berkovits, M.~Guillen and L.~Mason,
``Supertwistor description of ambitwistor strings,''
JHEP 01, 020 (2020). [arXiv:1908.06899 [hep-th]].
}

\lref\guillengarciaparticle{
D.~Garc{\'\i}a Sep{\'u}lveda and M.~Guillen,
``A pure spinor twistor description of the $D = 10$ superparticle,''
JHEP 08, 130 (2020). [arXiv:2006.06023 [hep-th]].
}

\lref\guillengarciaambitwistor{
D.~G.~Sep{\'u}lveda and M.~Guillen,
``A Pure Spinor Twistor Description of Ambitwistor Strings,'' [arXiv:2006.06025 [hep-th]].
}

\font\mbb=msbm10 
\newfam\bbb
\textfont\bbb=\mbb

\def\startcenter{%
  \par
  \begingroup
  \leftskip=0pt plus 1fil
  \rightskip=\leftskip
  \parindent=0pt
  \parfillskip=0pt
}
\def\stopcenter{%
  \par
  \endgroup
}

\listtoc
\writetoc
\filbreak

\newsec Introduction

10D super-Yang-Mills \tenDsYM\ and 11D supergravity \refs{\Nahm, \Scherk}\ are maximally supersymmetric gauge theories describing massless spin-1 and spin-2 particles, respectively. These theories arise as low-energy limits of superstring theory \refs{\GS, \GSsusy, \GSYMsugra, \psparticle, \ICTP, \superpoincarequantization} and M-theory, providing essential building blocks for understanding the unification of forces and quantum gravity in higher-dimensional frameworks.

\medskip
The physical states of 10D super-Yang-Mills can be easily described via light-cone gauge quantization of the 10D Brink-Schwarz superparticle \refs{\brinkschwarz, \tenLCG}. However, the application of this formalism to scattering amplitude computations is highly nontrivial and inherently constrained by the lack of manifest Lorentz invariance. A more simple and elegant quantization of the superparticle which preserves super-Poincar\'e covariance makes use of pure spinor variables \psparticle. This formulation has been shown to elegantly reproduce the physical states of the antifield description of 10D super-Yang-Mills \mafraoli. Owing to its manifest supersymmetry, the pure spinor formalism provides a powerful framework for computing their corresponding scattering amplitudes \refs{\stieberger, \mafratwo, \mafrathree, \Berkovitstwoloops, \BerkovitsMafratwoloops, \NMPS, \maxmaor}.

\medskip
\noindent These physical interactions are obtained from pure spinor correlation functions of vertex operators carrying ghost numbers one and zero. The ghost number one operator $U^{(1)}$ can be derived by perturbing the BRST charge $Q$ in the presence of an external background, while the ghost number zero operator $U^{(0)}$ arises from the corresponding variation of the Hamiltonian $H$. Conservation of the BRST charge $\{Q, H\} = 0$, implies a set of consistency relations among these vertex operators. In particular, linearized or single-particle vertex operators must satisfy the descent relation $[Q, U^{(0)}] = \partial_{\tau}U^{(1)}$. A comprehensive account of this approach at tree-level can be found in our companion paper \tendimensions.

\medskip
An analogous situation arises in 11D. The light-cone gauge quantization of the superparticle can be shown to reproduce the physical degrees of freedom of 11D supergravity \refs{\elevenpart}. Once again, the absence of manifest supersymmetry and Lorentz covariance renders this approach impractical for computing scattering amplitudes. This limitation is overcome by introducing pure spinor variables \refs{\pssupermembrane, \pssugra, \cederwallequations, \maxthesis}, which enable a covariant and supersymmetric quantization scheme suitable for amplitude calculations \refs{\maximalloopcederwall, \Karlsson, \elevendsuperspaceexpansion}.

\medskip
\noindent Unlike in 10D, the 11D pure spinor worldline formalism remains incomplete, owing to the current lack of knowledge regarding its foundational components. For instance, the existence of the ghost number zero vertex operator has been shown to be inconsistent with the equations of motion of 11D supergravity in the minimal setting \refs{\maxmasoncasaliberkovits, \maxnotesworldline}. Likewise, a concrete prescription for computing amplitudes has not yet been completely defined.

\medskip
In this work we address the first of these challenges. The obstruction identified in \maxmasoncasaliberkovits\ within the minimal formalism is circumvented by employing the non-minimal pure spinor superparticle. The inclusion of non-minimal variables allows for the construction of negative ghost number operators, referred to as physical operators, which have previously played a key role in significantly simplifying the expression for the b-ghost \refs{\berkovitsguillenbghost,\tamingbghost}. Guided by the principles outlined in \refs{\maxnotesworldline,\tamingbghost,\psborninfeld}, we define new physical operators consistent with 11D supergravity, that exhibit quadratic and cubic dependence on the worldline vector fields. We compute the action of a subset of these operators on the ghost number three vertex operator, and verify that they reproduce the associated superfields. As a key result, we use a small collection of these physical operators to construct a compact expression for the ghost number zero vertex operator. We demonstrate that this expression satisfies the descent relation with the 11D ghost number one operator, and yields the two-particle superfield introduced in \tamingbghost\ when acting on the ghost number three vertex.

\medskip
This paper is organized as follows: Section 2 reviews the minimal pure spinor superparticle in 11D, as well as the equations of motion of linearized 11D supergravity. Section 3 introduces the non-minimal pure spinor variables, and discusses the construction and properties of both linear and non-linear physical operators. Section 4 presents the ghost number zero vertex operator, and demonstrates its consistency with the standard descent relation and previously known results. Finally, Section 5 concludes with a discussion of our findings and potential future directions. Appendix A contains the calculation of the non-trivial commutation relations satisfied by certain linear physical operators. Explicit verifications of the properties of the non-linear physical operators are deferred to Appendices B and C, while Appendix D presents explicit computations of the action of selected non-linear physical operators.

\seclab\secone

\noindent

\newsec 11D Minimal Pure Spinor Superparticle

\seclab\sectwo

 The 11D pure spinor superparticle action in a curved background is given by \refs{\pssupermembrane,\pselevenparticle}
\eqnn \elevendpsaction
$$ \eqalignno{
S &= \int d\tau [P_M \partial_{\tau}Z^{M} + w_{\alpha}(\partial_{\tau}\lambda^{\alpha} + \partial_{\tau}Z^{M}\lambda^{\beta}\Omega_{M,\beta}{}^{\alpha}) - \half P^2] & \elevendpsaction
}
$$
We are using capital letters from the beginning/middle of the Latin alphabet to denote tangent/curved superspace indices, and lowercase letters from the beginning (middle) of the Latin/Greek alphabet to indicate tangent (curved) space vector/spinor indices. The variables $Z^{M} = (X^m, \theta^{\mu})$ are the usual 11D superspace coordinates, and $P_{M} = (P_m, p_{\mu})$ are their respective conjugate momenta. The bosonic spinor $\lambda^{\alpha}$ satisfies the pure spinor constraint in 11D, i.e. $\lambda\gamma^a \lambda = 0$. Due to this, its corresponding conjugate momentum $w_{\alpha}$ is only defined up to the gauge transformation $\delta w_{\alpha} = (\gamma^{a}\lambda)_{\alpha}\rho_a$, for any vector $\rho_{a}$. Since they possess wrong statistics, they will be called ghost variables, and be defined to carry ghost numbers 1 and -1, respectively. The superfields $E^{A} = \partial_{\tau}Z^{M}E_{M}{}^{A}$ and $\Omega_{A}{}^{B} = \partial_{\tau}Z^{M}\Omega_{M,A}{}^{B}$ are respectively the 11D supergravity vielbein and spin-connection. The 11D gamma matrices will be represented by $(\gamma^{a})_{\alpha\beta}$, $(\gamma^{a})^{\alpha\beta}$, and they satisfy the Clifford algebra: $(\gamma^{a})_{\alpha\beta}(\gamma^{b})^{\beta\delta} + (\gamma^{b})_{\alpha\beta}(\gamma^{a})^{\beta\delta} = 2\eta^{ab}\delta_{\alpha}^{\delta}$. In addition, the antisymmetric charge conjugation matrix $C_{\alpha\beta}$ and its inverse $C^{\alpha\beta}$, which obey the relation $C_{\alpha\beta}C^{\beta\delta} = \delta_{\alpha}^{\delta}$, will be used to raise and lower spinor indices, so that $(\gamma^{a})^{\alpha\beta} = C^{\alpha\epsilon}C^{\beta\delta}(\gamma^{a})_{\epsilon\delta}$, etc.

\medskip
The flat version of \elevendpsaction\ with BRST charge $Q_{0} = \lambda^{\alpha} d_{\alpha}$, where $d_{\alpha}$
is the Brink-Schwarz fermionic constraint \brinkschwarz, has been shown to reproduce all the fields and antifields of the Batalin-Vilkovisky description of linearized 11D supergravity \pssupermembrane. In particular, the antifield of the ghost-for-ghost-for-ghost of the gauge symmetry in 11D supergravity is located at the ghost number seven sector, and it represents the top cohomology of the BRST charge. As such, it has been proposed as the ideal candidate for defining manifestly supersymmetric correlation functions in 11D \pssupermembrane. 

\medskip

\noindent Likewise, the physical states of 11D supergravity can be found at the ghost number three cohomology. In order to illustrate this, let us first briefly review the superspace description of linearized 11D supergravity.

\subsec Linearized 11D supergravity

The 11D supergeometry is characterized by the torsion $T^{A} = {\cal D}E^{A}$, and curvature $R_{A}{}^{B} = {\cal D}\Omega_{A}{}^{B}$, with ${\cal D} = E^{A}\nabla_{A}$ being the super-covariant derivative. They satisfy the super-Bianchi identities
\eqnn \elevendgeometry
$$
\eqalignno{
{\cal D}T^{A} = E^{B}R_{B}{}^{A} \ \ &, \ \ {\cal D}R_{A}{}^{B} = 0  & \elevendgeometry\cr
}
$$
These equations in turn imply the familiar relations
\eqnn \torsiondef
\eqnn \curvaturedef
$$
\eqalignno{
[\nabla_{A}, \nabla_{B}\} &= - T_{AB}{}^{C}\nabla_{C} - 2\Omega_{[AB\}}{}^{C}\nabla_{C} , & \torsiondef\cr
R_{AB,C}{}^{D} &= 2\nabla_{[A}\Omega_{B\}C}{}^{D} + T_{AB}{}^{F}\Omega_{FC}{}^{D} + \Omega_{[AB\}}{}^{F}\Omega_{FC}{}^{D} & \curvaturedef
}
$$
where $[ \, ,\,\}$ stands for a graded commutator. The 3-form gauge field of 11D supergravity can be promoted to the 3-form superfield $F = E^{C}E^{B}E^{A}F_{ABC}$, defined up to the gauge transformation $\delta F = d L$, for any 2-form superfield $L$. The respective field strength is defined as $G = dF$, and it trivially satisfies $d G = 0$. The dynamical equations of linearized 11D supergravity are then found after expanding the covariant derivative $\nabla_{A} = E_{A}{}^{M}\partial_{M}$ up to first-order corrections, namely
\eqnn \introducingh
$$
\eqalignno{
\nabla_{A} &= D_{A} - h_{A}{}^{B}D_{B}& \introducingh 
}
$$
where $D_{A} = \hat{E}_{A}{}^{M}\partial_{M}$, $h_{A}{}^{B} = \hat{E}_{A}{}^{M}E_{M}^{(1)B} = -E^{(1)M}_{A}\hat{E}_{M}{}^{B}$, ($\hat{E}_{A}{}^{M}$, $\hat{E}_{M}{}^{B}$) are the flat-space values of the vielbeins, and ($E_{A}^{(1)M}$, $E_{M}^{(1)A}$) are their corresponding first-order perturbations. Furthermore, one should impose the so-called dynamical contraints $T_{\alpha\beta}{}^{a} = (\gamma^{a})_{\alpha\beta}$, $G_{\alpha\beta ab} = (\gamma_{ab})_{\alpha\beta}$, as well as the conventional constraints $T_{\alpha\beta}{}^{\delta} = T_{a\alpha}{}^{c} = T_{ab}{}^{c} = G_{\alpha\beta\delta\epsilon} = G_{a\alpha\beta\delta} = G_{abc\alpha} = 0$. After substituting \introducingh\ into \torsiondef, one obtains the following set of equations of motion \refs{\maxthesis,\maxmasoncasaliberkovits}
\eqnn \eomone
\eqnn \eomtwo
\eqnn \eomthree
\eqnn \eomfour
\eqnn \eomfive
\eqnn \eomsix
$$
\eqalignno{
2D_{(\alpha}h_{\beta)}{}^{a} - 2h_{(\alpha}{}^{\delta}(\gamma^{a})_{\beta)\delta} + h_{b}{}^{a}(\gamma^{b})_{\alpha\beta} &= 0 & \eomone \cr
2D_{(\alpha}h_{\beta)}{}^{\delta} - 2\Omega_{(\alpha\beta)}{}^{\delta} + (\gamma^{a})_{\alpha\beta}h_{a}{}^{\delta} &= 0 & \eomtwo\cr
\partial_{a}h_{\alpha}{}^{\beta} - D_{\alpha}h_{a}{}^{\beta} - T_{a\alpha}{}^{\beta} - \Omega_{a\alpha}{}^{\beta} &= 0 & \eomthree\cr
\partial_{a}h_{\alpha}{}^{b} - D_{\alpha}h_{a}{}^{b} - h_{a}{}^{\beta}(\gamma^{b})_{\beta\alpha} + \Omega_{\alpha a}{}^{b} &= 0 & \eomfour\cr
\partial_{a}h_{b}{}^{\alpha} - \partial_{b}h_{a}{}^{\alpha} - T_{ab}{}^{\alpha} &= 0 & \eomfive\cr
\partial_{a}h_{b}{}^{c} - \partial_{b}h_{a}{}^{c} - 2\Omega_{{ab}}{}^{c} &= 0 & \eomsix 
}
$$
It is straightforward to check eqns. \eomone-\eomsix\ are invariant under the gauge transformations
\eqnn \gtfull
$$
\eqalignno{
\delta h_{\alpha}{}^{a} = D_{\alpha}\Lambda^{a} + (\gamma^{a})_{\alpha\beta}\Lambda^{\beta} \ \ , \ \
\delta h_{\alpha}{}^{\beta} &= D_{\alpha}\Lambda^{\beta} + \Lambda_{\alpha}{}^{\beta} \ \ , \ \
\delta \Omega_{\alpha\beta}{}^{\epsilon} = D_{\alpha}\Lambda_{\beta}{}^{\epsilon} \ , &  \cr
\delta h_{a}{}^{b} = \partial_{a}\Lambda^{b} + \Lambda_{a}{}^{b} \ \ , \ \
\delta h_{a}{}^{\beta} &= \partial_{a}\Lambda^{\beta}\ \ , \ \ \delta \Omega_{a\alpha}{}^{\beta} = \partial_{a}\Lambda_{\alpha}{}^{\beta}  \ , & \gtfull
}
$$
where $\Lambda^{a}$, $\Lambda^{\alpha}$, $\Lambda_{\alpha}{}^{\beta} = {1\over 4}(\gamma^{ab})_{\alpha}{}^{\beta}\Lambda_{ab}$ are arbitrary gauge parameters. 

\medskip
The equations of motion associated to the linearized components of $F$, i.e. $C_{ABC} = \hat{E}_{[C}{}^{P}\hat{E}_{B}{}^{N}\hat{E}_{A\}}{}^{M}F_{MNP}$, can be derived from the 4-form superfield $H$ defined as
\eqnn \tensorcapitalh
$$
\eqalignno{
H_{ABCD} &= \hat{E}_{[D}{}^{Q}\hat{E}_{C}{}^{P}\hat{E}_{B}{}^{N}\hat{E}_{A\}}{}^{M}G_{MNPQ} & \tensorcapitalh 
}
$$
which can equivalently be written as $H_{ABCD} = 4D_{[A}C_{BCD\}} + 6\hat{T}_{[AB}{}^{E}C_{ECD\}}$, where $\hat{T}^{A}$ is the flat-space valued torsion. 
Eqn. \tensorcapitalh\ in components reads
\eqnn \eomseven
\eqnn \eomeight
\eqnn \eomnine
\eqnn \eomten
$$
\eqalignno{
4D_{(\alpha}C_{\beta\delta\epsilon)} + 6(\gamma^{a})_{(\alpha\beta}C_{a\delta\epsilon)} &= 0 & \eomseven\cr
 \partial_{a}C_{\alpha\beta\delta} - 3D_{(\alpha}C_{a\beta\delta)} + 3(\gamma^{b})_{(\alpha\beta}C_{ba\delta)}  &= -3(\gamma_{ab})_{(\alpha\beta}h_{\delta)}{}^{b} & \eomeight \cr
2\partial_{[a}C_{b]\alpha\beta} + 2D_{(\alpha}C_{\beta) ab} + (\gamma^{c})_{\alpha\beta}C_{cab} &= 2(\gamma_{[b}{}^{c})_{\alpha\beta}h_{a]c} - 2(\gamma_{ab})_{(\alpha\delta}h_{\beta)}{}^{\delta} & \eomnine\cr
3\partial_{[a}C_{bc]\alpha} - D_{\alpha}C_{abc} & =  -3(\gamma_{[ab})_{\alpha\beta}h_{c]}{}^{\beta} & \eomten
}
$$
It is not hard to show that these equations are invariant under the gauge transformations
\eqnn \gtone
\eqnn \gttwo
\eqnn \gtthree
\eqnn \gtfour
$$
\eqalignno{
\delta C_{\alpha\beta\epsilon} &= D_{(\alpha}\Lambda_{\beta\epsilon)} + (\gamma^{a})_{(\alpha\beta}\Lambda_{a\epsilon)}  \ ,& \gtone\cr
\delta C_{a\alpha\epsilon} &= {1\over 3}\partial_{a}\Lambda_{\alpha\epsilon} + {2\over 3}D_{(\alpha}\Lambda_{\epsilon)a} + 
{1\over 3}(\gamma^{b})_{\alpha\epsilon}\Lambda_{ba} + (\gamma_{ab})_{\alpha\epsilon}\Lambda^{b}\ , & \gttwo\cr
\delta C_{ab\alpha} &= {2
\over 3}\partial_{[a}\Lambda_{b]\alpha} + {1\over 3}D_{\alpha}\Lambda_{ab} -(\gamma_{ab})_{\alpha\beta}\Lambda^{\beta} \ ,& \gtthree\cr
\delta C_{abc} &= \partial_{[a}\Lambda_{bc]}  \ .& \gtfour
}
$$

\subsec Ghost number three vertex operator

\subseclab \sectwotwo

The ghost number three sector, $U^{(3)} = \lambda^{\alpha}\lambda^{\beta}\lambda^{\delta}A_{\alpha\beta\delta}$, in the pure spinor BRST cohomology must satisfy the following physical state conditions
\eqnn \closed
\eqnn \exact
$$ \eqalignno{
Q_{0}U^{(3)} = 0  &\rightarrow D_{(\alpha}A_{\beta\delta\epsilon)} = (\gamma^{a})_{(\alpha\beta}A_{a\delta\epsilon)} & \closed \cr
\delta U^{(3)} = Q_{0}\Sigma &\rightarrow \delta A_{\alpha\beta\delta} = D_{(\alpha}\Sigma_{\beta\delta)} + (\gamma^{a})_{(\alpha\beta}\Xi_{\delta)a} & \exact
}
$$
where $\Sigma = \lambda^{\alpha}\lambda^{\beta}\Sigma_{\alpha\beta}$, and $\Sigma_{\alpha\beta}$, $\Xi_{\delta a}$ are arbitrary superfields. Notice that the second term in eqn. \exact\ leaves $U^{(3)}$ invariant due to the pure spinor constraint. If one makes the identifications $A_{\alpha\beta\delta} = C_{\alpha\beta\delta}$, $A_{a\delta\epsilon} = -{3 \over 2}C_{a\delta\epsilon}$, $\Sigma_{\alpha\beta} = \Lambda_{\alpha\beta}$, $\Xi_{\delta a} = \Lambda_{\delta a}$, then eqns. \closed, \exact\ are nothing but the linearized equations of motion of 11D supergravity displayed in \eomseven, \gtone. More explicitly, one can fix a particular gauge to show that $U^{(3)}$ possesses the following $\theta$-expansion (see \elevendsuperspaceexpansion\ for details):
\eqnn \psithetaexpansion
$$ \eqalignno{
U^{(3)} =& -{1\over 8}(\lambda\gamma^{a}\theta)(\lambda\gamma^{b}\theta)(\lambda\gamma^{c}\theta)c_{abc} - {3\over 8}(\lambda\gamma^{cb}\theta)(\lambda\gamma_{c}\theta)(\lambda\gamma^{a}\theta)\epsilon_{ab}  - {1\over 5}(\lambda\gamma^{a}\theta)(\lambda\gamma^{b}\theta)(\lambda\gamma^{c}\theta)(\theta\gamma_{bc}\psi_{a})\cr
& + {1\over 5}(\lambda\gamma_{b}\theta)(\lambda\gamma^{a}\theta)(\lambda\gamma^{bc}\theta)(\theta\gamma_{c}\psi_{a}) + O(\theta^5) & \psithetaexpansion
}
$$
where $c_{abc}$, $\epsilon_{ab}$, $\psi^a_{\alpha}$ are respectively the 3-form, graviton and gravitino states of 11D supergravity.


\subsec Ghost number one vertex operator

\subseclab \sectwothree

In \maxmasoncasaliberkovits, a new vertex operator carrying ghost number one was found in the BRST cohomology of $Q_{0}$. This operator was constructed from the linear perturbation of the BRST charge induced by the coupling of the superparticle to a curved background. Explicitly, the BRST charge associated to the action \elevendpsaction\ reads
\eqnn \curvedbrstcharge
$$\eqalignno{
Q_{c} &= \lambda^{\alpha}E_{\alpha}{}^{M}(P_{M} + \Omega_{M,\alpha}{}^{\beta}w_{\beta}) &\curvedbrstcharge
}
$$
The linear deformation of $Q$ then defines the ghost number one vertex operator to be
\eqnn \ghostnumberonevo
$$\eqalignno{
U^{(1)} &= \lambda^{\alpha}(P_{a}h_{\alpha}{}^{a} + d_{\beta}h_{\alpha}{}^{\beta} - \Omega_{\alpha\,\beta}{}^{\delta}\lambda^{\beta}w_{\delta}) &\ghostnumberonevo
}
$$
where $\Omega_{A,B}{}^{C} = \hat{E}_{A}{}^{M}\Omega^{(1)}_{M,B}{}^{C}$, with $\Omega^{(1)}_{M,B}{}^{C}$ representing the linear contribution of the spin-connection. As a check, one can use eqns. \eomone, \eomtwo\ and (see \maxthesis\ for details)
\eqnn \randtforelevendsugraone
$$
\eqalignno{
R_{(\alpha\beta,\delta)}{}^{\epsilon} + (\gamma^{a})_{(\alpha\beta}T_{a\delta)}{}^{
\epsilon} &= 0  \ ,& \randtforelevendsugraone
}
$$
to readily show that $Q_{0}$ annihilates $U^{(1)}$. Likewise, the use of eqns. \gtfull\ allows one to show that the gauge variation of $U^{(1)}$ is BRST-exact.

\subsec Ghost number zero vertex operator: An issue

\subseclab \sectwofour

The computation of generic N-point correlators using the 11D pure spinor measure described above ultimately requires the insertion of ghost number zero vertex operators. These operators form an essential ingredient for evaluating scattering amplitudes with an arbitrary number of external states in 11D supergravity, since the total ghost number must be saturated by the pure spinor measure. For example, the prescription for the N-point function in the supermembrane formalism requires the insertion of $N-2$ ghost number zero vertex operators \pssupermembrane, while in the superparticle context the current proposal for computing N-point scattering amplitudes calls for the insertion of $N-3$ such operators \refs{\Grassi,\amplitudeeleven}.

\medskip 
\noindent In the latter framework, the corresponding operator is expected to satisfy the standard descent relation $\{Q_{0}, U^{(0)}\} = \partial_{\tau}U^{(1)}$. In \maxmasoncasaliberkovits, the authors proved that such a relation is inconsistent with 11D supergravity. This immediately follows from the fact that $T_{a\alpha}{}^{\beta}$ in the conventional description of 11D supergravity \brinkhowe, cannot be written as $T_{a\alpha}{}^{\beta} = (\gamma_{a})_{\alpha\delta}\cal{P}^{\delta\beta}$, for some $\cal{P}^{\delta\beta}$. This incompatibility was associated to the assumption of the existence of this vertex in the minimal pure spinor framework. In the next section, we will confirm this conjecture by solving the problem in the non-minimal pure spinor setup.

\medskip
It is worth mentioning that a ghost number two vertex operator satisfying a descent relation with $U^{(3)}$, i.e. $\{Q_{0}, U^{(2)}\} = \partial_{\tau}U^{(3)}$, has been constructed in \refs{\maxnotesworldline,\tamingbghost}. However, since such an operator will not play any relevant role in our present study, we will omit any discussion on this.





\newsec 11D Non-minimal Pure Spinor Superpaticle

\seclab\secthree

As in 10D, it is possible to extend the model \elevendpsaction\ to its so-called non-minimal version, which consists of adding a couple of conjugate variables $(\bar{\lambda}_{\alpha}, \bar{w}^{\alpha})$, $(r_{\alpha}, s^{\alpha})$ to the action \elevendpsaction, subject to the constraints $\bar\lambda \gamma^{a}\bar\lambda = \bar{\lambda}\gamma^{a}r = 0$. The non-minimal pure spinor superparticle action in a flat background is then given by
\eqnn \nonminimalpssuperparticle
$$
\eqalignno{
S &= \int d\tau [P_{m}\partial_{\tau}X^{m} + p_{\alpha}\partial_{\tau}\theta^{\alpha} + w_{\alpha}\partial_{\tau}\lambda^{\alpha} + \bar{w}^{\alpha}\partial_{\tau}\bar{\lambda}_{\alpha} + s^{\alpha}\partial_{\tau}r_{\alpha} - \half P^2 ] & \nonminimalpssuperparticle
}
$$
The non-minimal BRST charge is topologically modified to the form  
\eqnn \nonminimalbrstcharge
$$
\eqalignno{
Q &= Q_{0} + r_{\alpha}\bar{w}^{\alpha} & \nonminimalbrstcharge
}
$$
so that the cohomology of Q is equivalent to that of $Q_{0}$. The action \nonminimalpssuperparticle\ is invariant under the global symmetry $J = w_{\alpha}\lambda^{\alpha} -\bar{w}^{\alpha}\bar{\lambda}_{\alpha}$, referred to as the non-minimal ghost number charge. In this manner, $\bar{\lambda}_{\alpha}$ and $r_{\alpha}$ carry ghost numbers -1 and 0, respectively.

\medskip
The advantage of the non-minimal framework over the minimal one, is that the former admits the construction of Lorentz covariant negative ghost number operators. Indeed, the authors of \maximalloopcederwall\ used the non-minimal formalism in order to define a b-ghost satisfying the standard relation $\{Q, b\} = P^{2}$.

\medskip
More recently, one of the authors made use of the non-minimal variables to define a set of negative ghost number operators with simple analytical properties \refs{\maxnotesworldline,\tamingbghost}. These operators were then used for notably simplifying the expression of the b-ghost, and so to make its algebraic manipulations much more tractable and efficient. Next, we briefly review the construction of these operators and introduce a new set of these, which will turn out to be fundamental for the construction of the ghost number zero vertex operator.

\subsec Linear physical operators

\subseclab \secthreeone

The 11D physical operators studied in \tamingbghost\ were defined as linear operators acting on the ghost number three vertex operator $U^{(3)}$, so that up to extra terms (shift symmetry and BRST-exact terms) it reproduces the superfield which it is associated to, in a way that is consistent with the equations of motion \eomone, \eomtwo, \eomseven, \eomeight. More concretely, after contracting these equations with a number of pure spinors, one finds
\eqnn \drchatalpha
\eqnn \drchata
\eqnn \drphihata
\eqnn \drphihatalpha
$$ \eqalignno{
[Q, {\bf C}_{\alpha}] &= -{1\over 3}d_{\alpha} - (\gamma^a \lambda)_{\alpha}{\bf C}_{a}  & \drchatalpha \cr
\{Q, {\bf C}_{a}\} &= {1\over 3}P_{a} -(\lambda\gamma^{ab}\lambda){\bf \Phi}_{b} 
 & \drchata \cr
[Q, {\bf \Phi}^{a}] &= (\lambda\gamma^a {\bf \Phi}) & \drphihata \cr
\{Q, {\bf \Phi}^{\alpha}\} &= {1\over 4}(\lambda\gamma^{ab})^{\alpha}{\bf \Omega}_{ab} & \drphihatalpha \cr
\vdots
}
$$
where we are using bold capital letters to represent the corresponding physical operators (see Appendix A of \tamingbghost\ for details). The ellipsis below eqn. \drphihatalpha\ represents operator relations defining higher-order physical operators. These operators are not relevant for our purposes, and we therefore omit their discussion. One can check that this system of equations is solved by the following expressions
\eqnn \chatalpha
\eqnn \chata
\eqnn \phihata
\eqnn \phihatalpha
$$
\eqalignno{
{\bf C}_{\alpha} &= {1\over 3}K_{\alpha}{}^{\beta}w_{\beta}  & \chatalpha\cr
{\bf C}^{a} &= {1 \over \eta} (\lambda\gamma^{abc})^{\alpha}(\bar{\lambda}\gamma_{bc}\bar{\lambda})\bigg[{1 \over 3}d_{\alpha} + [Q, {\bf C}_{\alpha}]\bigg] & \chata \cr 
{\bf \Phi}^{a} &= -{2 \over \eta} (\bar{\lambda}\gamma^{ab}\bar{\lambda})\bigg[{1\over 3}P_{b} - \{Q, {\bf C}_{b}\}\bigg] & \phihata \cr
{\bf \Phi}^{\alpha} &= {2\over \eta}\xi_{a}^{\alpha}[Q, {\bf \Phi}^{a}] & \phihatalpha
}
$$
where $\eta = (\lambda\gamma_{ab}\lambda)(\bar{\lambda}\gamma^{ab}\bar{\lambda})$, and $K_{\alpha}{}^{\beta}$ is the 11D pure spinor projector given by \tamingbghost
\eqnn \psprojectorminecompact
$$
\eqalignno{
K_{\alpha}{}^{\beta} &= \delta_{\alpha}^{\beta} + {1\over \eta}(\lambda\gamma^{abc})^{\beta}(\bar{\lambda}\gamma_{bc}\bar{\lambda})(\lambda\gamma_{a})_{\alpha} & \psprojectorminecompact
}
$$
It is also straightforward to verify that the operators \chatalpha-\phihatalpha\ satisfy the constraints
\eqnn \constraintsforop
$$
\eqalignno{
\xi_{a}^{\alpha}{\bf C}_{\alpha} &= 0, \ \ \ \  \ (\bar{\lambda}\gamma^{ab}\bar{\lambda}){\bf C}_{a} = 0, \ \ \ \ \ (\bar{\lambda}\gamma^{a})_{\alpha}{\bf \Phi}_{a} = 0, \ \ \ \ \ R_{\alpha}{}^{\beta}{\bf \Phi}^{\alpha} = 0 & \constraintsforop
}
$$
with $\xi_{a}^{\beta}$ and $R_{\alpha}{}^{\beta}$ taking the explicit forms
\eqnn \xiaalpha
\eqnn \ralphabeta
$$
\eqalignno{
\xi_{a}^{\beta} =& \half (\gamma_{abc})^{\beta\delta}\lambda_{\delta}(\bar{\lambda}\gamma^{bc}\bar{\lambda}) &  \xiaalpha\cr
R_{\alpha}{}^{\beta}=& \bigg[-{1\over 2}(\lambda\gamma^{b})_{\alpha}(\lambda\gamma^{c})^{\beta} - {1\over 4}(\lambda\gamma^{bk}\lambda)(\gamma^{c}\gamma^{k})_{\alpha}{}^{\beta} + {1\over 2}(\lambda\gamma^{bk})_{\alpha}(\lambda\gamma^{ck})^{\beta} - {1\over 2}(\lambda\gamma^{bc})_{\alpha}\lambda^{\beta}\bigg](\bar{\lambda}\gamma_{bc}\bar{\lambda}) &  \cr
&  &\ralphabeta
}
$$
These objects have been used to construct a ghost number -2 operator which maps the cohomology of the ghost number three vertex operator into that of the ghost number one vertex operator \maxnotesworldline. They obey the useful relations
\eqnn \usefulrelations
$$
\eqalignno{
(\lambda\gamma_{ac}\lambda)(\bar{\lambda}\gamma^{bc}\bar{\lambda}) &= {1\over 2}\delta_{a}^{b}\eta - (\lambda\gamma^{b}\xi_{a}), \ \ \ \ \ (\lambda\gamma^{a})_{\alpha}\xi_{a}^{\beta} = \half \delta_{\alpha}^{\beta}\eta + R_{\alpha}{}^{\beta}, & \usefulrelations
}
$$


\medskip
\noindent The explicit form of these operators then reads
\eqnn \chatalphaexp
\eqnn \chataexp
\eqnn \phihataexp
\eqnn \phihatalphaexp
$$
\eqalignno{
{\bf C}_{\alpha} =& {w_{\alpha} \over 3} + {1\over 3\eta}(\lambda\gamma^{abc}w)(\bar{\lambda}\gamma_{bc}\bar{\lambda})(\lambda\gamma_{a})_{\alpha} & \chatalphaexp \cr
{\bf C}_{a} =& {1\over 3\eta}(\bar{\lambda}\gamma^{bc}\bar{\lambda})(\lambda\gamma_{abc}d) - {2\over 3\eta}(\bar{\lambda}\gamma^{bc}r)(\lambda\gamma_{abc}w) + {2\over 3\eta^2}\phi (\bar{\lambda}\gamma^{bc}\bar{\lambda})(\lambda\gamma_{abc}w) & \cr
& +{4\over 3 \eta^{2}}(\lambda\gamma_{ac}\lambda)(\bar{\lambda}\gamma^{bc}\bar{\lambda})(\bar{\lambda}\gamma^{de}r)(\lambda\gamma_{bde}w)  & \chataexp \cr
{\bf \Phi}^{a} =& -{2 \over 3}\bigg[{1 \over \eta}(\bar{\lambda}\gamma^{ab}\bar{\lambda})P_{b} - {2 \over \eta^{2}}(\bar{\lambda}\gamma^{ab}\bar{\lambda})(\bar{\lambda}\gamma^{cd}r)(\lambda\gamma_{bcd}d) - \{s, {2\over \eta^2}(\bar{\lambda}\gamma^{ab}\bar{\lambda})(\bar{\lambda}\gamma^{cd}r)\}(\lambda\gamma_{bcd}w) & \cr
& - {8 \over \eta^3}(\lambda\gamma^{a}\xi_{b})(\bar{\lambda}\gamma^{cb}r)(\bar{\lambda}\gamma^{de}r)(\lambda\gamma_{cde}w)\bigg] & \phihataexp \cr
{\bf \Phi}^{\alpha} =& -{8 \over 3}\xi_{a}^{\alpha}\bigg[{1\over \eta^{2}}(\bar{\lambda}\gamma^{ab}r)P_{b} - {2 \over \eta^{3}}(\bar{\lambda}\gamma^{ab}r)(\bar{\lambda}\gamma^{cd}r)(\lambda\gamma_{bcd}d) & \cr
& - \bigg({8 \over \eta^{4}} (\bar{\lambda}\gamma^{ab}r)(\lambda\gamma_{cb}\lambda)(\bar{\lambda}\gamma^{cd}r)(\bar{\lambda}\gamma^{ef}r) + {8 \over \eta^{4}}\phi(\bar{\lambda}\gamma^{ab}r) (\bar{\lambda}\gamma^{cd}r)\bigg)(\lambda\gamma_{bcd}w)\bigg] & \cr
&  & \phihatalphaexp 
}
$$
where $\phi = (\lambda\gamma^{ab}\lambda)(\bar{\lambda}\gamma_{ab}r)$. Eqns. \chatalphaexp-\phihatalphaexp\ were presented for the first time in \tamingbghost, and used to compute the action of the 11D b-ghost on $U^{(3)}$. Note that eqns. \phihataexp, \phihatalphaexp\ appear to slightly differ from those in \tamingbghost. However, the use of the identities
\eqnn \identitiesofxi
$$
\eqalignno{
(\bar{\lambda}\gamma^{ab}r)\xi_{a}^{\alpha}\xi_{b}^{\beta} & = 0, \ \ \ \ \ \xi_{a}^{\alpha}(\lambda\gamma^{a}\xi_{b}) = {\eta\over 2}\xi_{b}^{\alpha}, &\identitiesofxi
}
$$
allows one to readily show that they are actually equivalent to each other.

\medskip
 One can continue with this procedure, and define new linear operators compatible with 11D supergravity. To this end, let us recall some of the supergravity equations of motion (see \maxthesis\ for details)
\eqnn \rwithtaab
\eqnn \rawithdtaab
$$
\eqalignno{
R_{\alpha\beta,a}{}^{b} + 2(\gamma^{b})_{(\alpha\delta}T_{a\beta)}{}^{\delta} & = 0 & \rwithtaab \cr
R_{a(\alpha,\beta)}{}^{\delta} - D_{(\alpha}T_{a\beta)}{}^{\delta} - {1\over 2}(\gamma^{b})_{\alpha\beta}T_{ab}{}^{\delta} & = 0 & \rawithdtaab
}
$$
These formulae then give rise to the relations
\eqnn \morephysoptab
\eqnn \morephysopracd
$$
\eqalignno{
[Q, {\bf \Omega}_{a}{}^{b}] &= - (\lambda\gamma^{b})_{\delta}{\bf T}_{a}{}^{\delta} & \morephysoptab\cr
\{Q, {\bf T}_{a}{}^{\delta}\} &= {1\over 4}(\lambda\gamma^{cd})^{\delta}{\bf R}_{a,cd} & \morephysopracd
}
$$
where we defined ${\bf T}_{a}{}^{\delta} = \lambda^{\alpha}{\bf T}_{a\alpha}{}^{\delta}$, ${\bf R}_{a,cd} = \lambda^{\alpha}{\bf R}_{a\alpha,cd}$.

\medskip
\noindent In order to solve eqns. \drphihatalpha, \morephysoptab\ and \morephysopracd, one should introduce the tensor $\rho^{ab}_{\alpha}$ defined as
\eqnn \rhotensor
$$
\eqalignno{
\rho^{cd}_{\alpha} &= \lambda_{\alpha}(\bar{\lambda}\gamma^{cd}\bar{\lambda})+ 2(\gamma^{[ck}\lambda)_{\alpha}(\bar{\lambda}\gamma^{d]}{}_{k}\bar{\lambda}) + {1\over 4}(\gamma^{cdrs}\lambda)_{\alpha}(\bar{\lambda}\gamma_{rs}\bar{\lambda}) & \rhotensor
}
$$
which satisfies the useful identity
\eqnn \idforrhoabalpha
$$
\eqalignno{
(\lambda\gamma^{ab})^{\alpha}\rho_{cd\alpha} &= -{3\over 2}\delta_{cd}^{ab}\eta -{1\over 2} (\lambda\gamma_{cd}\lambda)(\bar{\lambda}\gamma^{ab}\bar{\lambda}) + 4(\lambda\gamma^{\underline{b}}{}_{[c}\lambda)(\bar{\lambda}\gamma_{d]}{}^{\underline{a}}\bar{\lambda})  -4(\lambda\gamma_{[cs}\lambda)(\bar{\lambda}\gamma^{\underline{b}s}\bar{\lambda})\delta^{\underline{a}}_{d]}& \cr
& + {1\over 4}\bigg[(\lambda\gamma^{ab}{}_{cdrs}\lambda)(\bar{\lambda}\gamma^{rs}\bar{\lambda})+ 4(\lambda\gamma^{ab}\lambda)(\bar{\lambda}\gamma_{cd}\bar{\lambda}) - 16(\lambda\gamma^{\underline{b}k}\lambda)(\bar{\lambda}\gamma_{[dk}\bar{\lambda})\delta^{\underline{a}}_{c]} + 4\delta^{ab}_{cd}\eta\bigg] & \cr 
& & \idforrhoabalpha
}
$$
In this manner, one learns that
\eqnn \solforomeab
\eqnn \solfortadelta
\eqnn \solforracd
$$
\eqalignno{
{\bf \Omega}_{cd} &= -{8\over 3\eta}\rho_{cd\alpha}\{Q, {\bf \Phi}^{\alpha}\} & \solforomeab \cr
{\bf T}_{a}{}^{\alpha} &= -{2\over \eta}\xi_{b}^{\alpha}[Q, {\bf \Omega}_{a}{}^{b}] & \solfortadelta \cr
{\bf R}_{a,cd} & = -{8 \over 3\eta}\rho_{cd\delta}\{Q, {\bf T}_{a}{}^{\delta}\} & \solforracd
}
$$
The use of standard pure spinor identities, and the relation $R_{\alpha}{}^{\beta} = {1\over 3}\rho_{\alpha}^{ab}(\lambda\gamma_{ab})^{\beta} - {1\over 6}(\lambda\gamma^{ab})_{\alpha}\lambda^{\beta}(\bar{\lambda}\gamma_{ab}\bar{\lambda})$, enables one to show that these operators obey the following constraints
\eqnn \idfornewop
$$
\eqalignno{
(\bar{\lambda}\gamma^{ab}\bar{\lambda}){\bf \Omega}_{bc} &= 0,\ \ \ \ \ \ \ \ (\bar{\lambda}\gamma^{ab}\bar{\lambda}){\bf R}_{c,bd} = 0,\ \ \ \ \ \ \ \ R_{\delta}{}^{\alpha}{\bf T}_{a}{}^{\delta} = 0, & \cr 
{\cal R}_{ab}{}^{cdef}&(\bar{\lambda}\gamma_{cd}\bar{\lambda}){\bf \Omega}_{ef} = 0, \ \ \ \ {\cal R}_{ab}{}^{cdef}(\bar{\lambda}\gamma_{cd}\bar{\lambda}){\bf R}_{g,ef} = 0,  & \idfornewop
}
$$
where ${\cal R}_{ab}{}^{cdef} = \delta^{[c}_{a}\delta^{d}_{b}(\lambda\gamma^{ef]}\lambda) + {1\over 24}(\lambda\gamma_{ab}{}^{cdef}\lambda)$. Explicitly, these operators are found to be
\eqnn \omegacdexp
\eqnn \taalphaexp
\eqnn \racdexp
$$
\eqalignno{
{\bf \Omega}_{ab} &= -{32\over \eta}\bigg[({\bf \Phi}\gamma_{[ak}\lambda)(\bar{\lambda}\gamma_{b]}{}^{k}r) + {1\over 8}({\bf \Phi}\gamma_{abcd}\lambda)(\bar{\lambda}\gamma^{cd}r)\bigg]& \omegacdexp \cr
{\bf T}_{a}{}^{\alpha} & = {2\over \eta}(\gamma^{bcd}\lambda)^{\alpha}(\bar{\lambda}\gamma_{cd}r){\bf \Omega}_{ab} & \taalphaexp \cr
{\bf R}_{a,cd} &=  -{32 \over \eta}\bigg[{\bf T}_{a}{}^{\delta}(\gamma_{[ck}\lambda)_{\delta}(\bar{\lambda}\gamma_{d]}{}^{k}r) + {1\over 8} {\bf T}_{a}{}^{\delta}(\gamma_{cdrs}\lambda)_{\delta}(\bar{\lambda}\gamma^{rs}r)\bigg] & \racdexp
}
$$
where we used that $\rho_{\alpha}^{cd}\xi_{a}^{\alpha} = 0$. The formulae describing the linear physical operators presented in this section, i.e. eqns. \chatalphaexp, \chataexp, \phihataexp, \phihatalphaexp, \omegacdexp, \taalphaexp, \racdexp, have been explicitly written in terms of the non gauge-invariant object $w_{\alpha}$, instead of the usual gauge-invariant currents $N^{ab} = {1\over 2}(\lambda\gamma^{ab}w)$, $J = \lambda w$. This might naively lead one to conclude that these operators are gauge-dependent quantities. However, this is not the case, as it can easily be seen from the structure of our construction whose building block was ${\bf C}_{\alpha}$, a gauge-invariant object by definition. Indeed, as discussed in \tamingbghost, the physical operators ${\bf C}_{\alpha}$, ${\bf C}_{a}$, ${\bf \Phi}^{a}$, ${\bf \Phi}^{\alpha}$ can be shown to be expressible in terms of the currents $N^{ab}$, $J$. Since the operators ${\bf \Omega}_{ab}$, ${\bf T}_{a}{}^{\alpha}$, ${\bf R}_{a,bc}$ can be written in terms of the above, it is straightforward to see that they also admit manifestly gauge-invariant forms.

\subsec Non-linear physical operators

\subseclab \secthreetwo

The operators discussed in the previous section depend linearly on worldline vectors. As will be shown now, it is also possible to introduce operators with a non-linear dependence on these variables, and whose action on $U^{(3)}$ also results in physical superfields. In particular, this non-linearity will be crucial for defining physical operators giving rise to ordinary ghost number zero superfields, e.g. those describing the linearized 11D supergeometry.

\medskip
Let us start by introducing quadratic physical operators. The corresponding defining relations can be simply deduced from the equations of motion \eomone-\eomfour\ and \eomseven-\eomnine, after contracting them with one and two pure spinor variables, respectively. They read
\eqnn \morephysophalphab
\eqnn \morephysophalphabeta
\eqnn \morephysophabeta
\eqnn \morephysophab
\eqnn \morephysopcalphabeta
\eqnn \morephysopcalphab
\eqnn \morephysopcab
$$
\eqalignno{
[Q, {\bf h}_{\alpha}{}^{a}] & = - D_{\alpha}{\bf \Phi}^{a} + (\lambda\gamma^{a})_{\delta}{\bf h}_{\alpha}{}^{\delta} + (\gamma^{a}{\bf \Phi})_{\alpha} - (\lambda\gamma^{b})_{\alpha}{\bf h}_{b}{}^{a} & \morephysophalphab
\cr
\{Q, {\bf h}_{\alpha}{}^{\beta}\}&=  - D_{\alpha}{\bf \Phi}{}^{\beta} + {1\over 4}(\gamma^{ab})_{\alpha}{}^{\beta}{\bf \Omega}_{ab} + {1\over 4}(\lambda\gamma^{ab})^{\beta}{\bf \Omega}_{\alpha ab} - (\lambda\gamma^{a})_{\alpha}{\bf h}_{a}{}^{\beta}  & \morephysophalphabeta \cr
[Q , {\bf h}_{a}{}^{\beta}] &= \partial_{a}{\bf \Phi}^{\beta} - {\bf T}_{a}{}^{\beta} - {1\over 4}(\lambda\gamma^{cd})^{\beta}{\bf \Omega}_{acd} & \morephysophabeta \cr
\{Q , {\bf h}_{a}{}^{b}\} & = \partial_{a}{\bf \Phi}^{b} - (\lambda\gamma^{b})_{\beta}{\bf h}_{a}{}^{\beta} + {\bf \Omega}_{a}{}^{b} & \morephysophab \cr
[Q, {\bf C}_{\alpha\beta}] &= -D_{(\alpha}{\bf C}_{\beta)} - 2(\lambda\gamma^{a})_{(\alpha}{\bf C}_{a \beta)} - {1\over 2}(\gamma^{a})_{\alpha\beta}{\bf C}_{a} & \morephysopcalphabeta\cr
\{Q, {\bf C}_{\delta a}\} &= {1\over 2}(D_{\delta}{\bf C}_{a} - \partial_{a}{\bf C}_{\delta}) - (\lambda\gamma^{b})_{\delta}{\bf C}_{ba} + {1\over 2}(\lambda\gamma_{ab}\lambda){\bf h}_{\delta}{}^{b} - (\lambda\gamma_{ab})_{\delta}{\bf \Phi}^{b} & \morephysopcalphab \cr
[Q, {\bf C}_{ab}] &= -\partial_{[a}{\bf C}_{b]} + (\lambda\gamma_{[a}{}^{c}\lambda){\bf h}_{b]c} - (\lambda\gamma_{ab})_{\delta}{\bf \Phi}^{\delta} & \morephysopcab
}
$$
Remarkably, these equations are automatically solved after making the following identifications (see Appendix B for an explicit verification)
\eqnn \quadraticop
$$
\eqalignno{
{\bf C}_{\alpha\beta} = {3\over 2}{\bf C}_{(\alpha}\circ {\bf C}_{\beta)},& \ \ {\bf C}_{a \alpha} = {3\over 4}({\bf C}_{a}\circ {\bf C}_{\alpha} + {\bf C}_{\alpha}\circ {\bf C}_{a}) + {1\over 4\eta}(\lambda\gamma^{apq}\gamma^{b})_{\alpha}(\bar{\lambda}\gamma_{pq}\bar{\lambda}){\bf C}_{b}, \ \ {\bf C}_{ab} = -{3\over 2}{\bf C}_{[a} \circ {\bf C}_{b]}, &\cr
 \ \  {\bf h}_{\alpha}{}^{a} = {3\over 2}&({\bf C}_{\alpha} \circ {\bf \Phi}^{a} + {\bf \Phi}^{a}  \circ {\bf C}_{\alpha}) - {1\over \eta}(\gamma^{pq}\lambda)_{\alpha}(\bar{\lambda}\gamma_{pq}\bar{\lambda}){\bf \Phi}^{a} - {4\over \eta^{2}}(\bar{\lambda}\gamma^{pa}r)\xi_{p}^{\epsilon}(\gamma^{c})_{\epsilon\alpha}{\bf C}_{c}, & \cr
& \ \ \ \ \ \ \ \ \ \ \ \ \ \ \ {\bf h}_{a}{}^{b} = {3\over 2}({\bf C}_{a}\circ {\bf \Phi}^{b} + {\bf \Phi}^{b}\circ {\bf C}_{a}), &\cr
{\bf h}_{\alpha}{}^{\delta} = {3\over 2}\bigg[&{\bf C}_{\alpha}\circ {\bf \Phi}^{\delta} + {\bf \Phi}^{\delta}\circ {\bf C}_{\alpha} - {2\over 3\eta}\xi_{n}^{\delta}(\gamma^{n}{\bf \Phi}) - {8\over 3\eta^{2}}\xi_{f}^{\delta}(\bar{\lambda}\gamma^{ft}r)(\gamma^{mn}\lambda)_{\alpha}(\bar{\lambda}\gamma_{mn}\bar{\lambda}){\bf \Sigma}_{t} & \cr
& \ \ \ \ \ \ \ \ \ \ \ \ \ + {32\over 3\eta^{3}}[Q,\xi_{t}^{\delta}](\bar{\lambda}\gamma^{pt}r)\xi_{p}^{\delta}(\gamma^{s})_{\delta\alpha}{\bf C}_{s}\bigg] &\cr
& \vdots &\quadraticop
}
$$
where $\circ$ stands for operator composition, $[{\bf C}_{\alpha}, {\bf C}_{a},{\bf \Phi}^{a}, {\bf \Phi}^{\alpha}, {\bf \Omega}_{ab}]$ are the linear operator presented in section \secthreeone, and we used ${\bf \Phi}^{a} = (\bar{\lambda}\gamma^{ab}\bar{\lambda}){\bf \Sigma}_{b}$. The ellipsis in eqn. \quadraticop\ denotes higher-order operators which can be straightforwardly constructed by following the simple algebraic procedure spelled out in Appendix B. The presence of terms explicitly depending on powers of $\eta$ in some of the operators displayed above is a direct consequence of the non-trivial commutation relations between the physical operators and pure spinor variables. As will be seen in the next subsection, these expressions will turn out to be quite useful when computing the action of the quadratic physical operators on $U^{(3)}$.

\medskip
Analogously, one can define cubic physical operators by using eqns. \eomseven-\eomten, and contracting them with one single pure spinor variable. It is not difficult to show that they are constrained by the following relations
\eqnn \cubicopone
\eqnn \cubicoptwo
\eqnn \cubicopthree
\eqnn \cubicopfour
$$
\eqalignno{
[Q, {\bf C_{\alpha\beta\delta}}] &= - 3D_{(\alpha}{\bf C}_{\beta\delta)} - 3(\lambda\gamma^{a})_{(\alpha}{\bf C}_{a\beta\delta)} - 3(\gamma^{a})_{(\alpha\beta}{\bf C}_{a\delta)} & \cubicopone\cr
\{Q, {\bf C}_{a\beta\delta}\} & = \partial_{a}{\bf C}_{\beta\delta}
- 2D_{(\beta}{\bf C}_{a\delta)} + 2(\lambda\gamma^{b})_{(\beta}{\bf C}_{ba\delta)} + (\gamma^{b})_{\beta\delta}{\bf C}_{ba} + 2(\lambda\gamma_{ab})_{(\beta}{\bf h}_{\delta)}{}^{b} + (\gamma_{ab})_{\beta\delta}{\bf \Phi}^{b} & \cr
& &\cubicoptwo\cr
[Q, {\bf C}_{\alpha ab}] &= -D_{\alpha}{\bf C}_{ab} - 2\partial_{[a}{\bf C}_{b]\alpha } - (\lambda\gamma^{c})_{\alpha}{\bf C}_{cab} - 2(\lambda\gamma_{[a c})_{\alpha}{\bf h}_{b]}{}^{c} - (\lambda\gamma_{ab})_{\delta}{\bf h}_{\alpha}{}^{\delta} - (\gamma_{ab})_{\alpha\delta}{\bf \Phi}^{\delta} &\cr 
& &\cubicopthree\cr
\{Q, {\bf C}_{abc}\} &= 3\partial_{[a}{\bf C}_{bc]} + 3(\lambda\gamma_{[ab})_{\beta}{\bf h}_{c]}{}^{\beta} & \cubicopfour
}
$$
As before, these equations become identities once one defines (see Appendix C for details)
\eqnn \cubicsolone
\eqnn \cubicsoltwo
$$
\eqalignno{
{\bf C}_{\alpha\beta\delta} &= {9\over 2}{\bf C}_{\alpha} \circ {\bf C}_{\beta} \circ {\bf C}_{\delta} + {1\over 4\eta}(\gamma^{a})_{(\alpha\beta}(\gamma_{pqa})_{\delta)}{}^{\kappa}(\bar{\lambda}\gamma^{pq}\bar{\lambda}){\bf C}_{\kappa} & \cubicsolone\cr
{\bf C}_{\alpha\beta a} &=  2{\bf C}_{\beta}\circ{\bf C}_{a\delta)} + {\bf C}_{a}\circ {\bf C}_{\beta\delta)} + {1 \over \eta}(\lambda\gamma^{pqa}\gamma^{r})_{\beta}(\bar{\lambda}\gamma_{pq}\bar{\lambda}){\bf C}_{r\delta)} &\cr
& + {1\over 2\eta}(\gamma^{b})_{\beta\delta)}[\xi_{b}^{\epsilon}, F_{a}]{\bf C}_{\epsilon} - {1\over 6\eta}(\gamma^{f})_{(\alpha\beta}\xi_{f}^{\epsilon}[Q, {1\over \eta}(\bar{\lambda}\gamma^{pq}\bar{\lambda})(\gamma_{pqa})_{\epsilon}{}^{\kappa}{\bf C}_{\kappa}]&\cr
& + {2\over 3\eta}(\gamma_{at}\lambda)_{\beta)}(\bar{\lambda}\gamma^{rt}\bar{\lambda})(\gamma_{pqr})_{\delta)}{}^{\kappa}[Q, {1\over \eta}(\bar{\lambda}\gamma^{pq}\bar{\lambda}){\bf C}_{\kappa}] &\cr
& - {2\over 9\eta}(\gamma^{f})_{(\beta\delta}\xi^{\epsilon}_{f}[Q,{1\over \eta^{2}}(\lambda\gamma^{cd}w)(\bar{\lambda}\gamma_{cd}\bar{\lambda})(\bar{\lambda}\gamma_{ag}\bar{\lambda})(\lambda\gamma^{g})_{\epsilon}] &\cubicsoltwo\cr 
& \vdots &
}
$$
where $F_{a}$ is defined from the equality $[Q,{\bf C}_{\alpha}] = (\gamma^{a})_{\alpha}F_{a}$, and eqn. \chataexp. In this manner, we have found a set of non-minimal operators carrying ghost number -1: [${\bf C}_{\alpha}$, ${\bf C}_{a}$], ghost number -2: [${\bf \Phi}^{a}$, ${\bf \Phi}^{\alpha}$, ${\bf \Omega}_{ab}$, ${\bf T}_{a}{}^{\alpha}$, ${\bf R}_{a,bc}$, ${\bf C}_{\alpha\beta}$, ${\bf C}_{\alpha a}$, ${\bf C}_{ab}$], and ghost number -3: [${\bf C}_{\alpha\beta\delta}$, ${\bf C}_{\alpha\beta a}$,  ${\bf C}_{\alpha bc}$,  ${\bf C}_{abc}$], which satisfy a number of operator equations determined by the dynamics of 11D supergravity, and reproduce pure spinor superfields after acting on the ghost number three vertex operator $U^{(3)}$. Next, we show how this explicitly works for various physical operators.

\subsec Physical operators and $U^{(3)}$

\subseclab \secthreethree

The action of some of the linear operators studied in section \secthreeone\ on $U^{(3)}$ has been explicitly computed in \tamingbghost. Let us briefly review how these manipulations are carried out, and then let us implement them on the newly defined operators.

\medskip
\noindent The formula \chatalphaexp\ immediately implies that
\eqnn \chatalphaonpsi
$$
\eqalignno{
{\bf C}_{\alpha}U^{(3)} &= C_{\alpha} + (\lambda\gamma^{a})_{\alpha}\rho_{a} & \chatalphaonpsi
}
$$
where $C_{\alpha} = \lambda^{\beta}\lambda^{\delta}C_{\alpha\beta\delta}$, and $\rho^{a} = {1\over \eta}(\lambda\gamma^{abc})^{\alpha}C_{\alpha}(\bar{\lambda}\gamma_{bc}\bar{\lambda})$. This equation in turn implies that the action of ${\bf C}^{a}$ on $U^{(3)}$ is given by
\eqnn \chataonpsi
$$
\eqalignno{
{\bf C}_{a}U^{(3)} &= C_{a} + (\lambda\gamma_{ac}\lambda)s^{c} - Q\rho_{a} & \chataonpsi
}
$$
where $C_{a} = \lambda^{\beta}\lambda^{\delta}C_{a\beta\delta}$, and $s^{b} = {2\over \eta}(\bar{\lambda}\gamma^{bc}\bar{\lambda})(C_{c}-Q\rho_{c})$. Similarly, the use of eqn. \chataonpsi\ allows one to state that 
\eqnn \phihataonpsi
$$
\eqalignno{
{\bf \Phi}^{a}U^{(3)} &= \Phi^{a} + (\lambda\gamma^{a}\kappa) - Qs^a & \phihataonpsi 
}
$$
where $\Phi^{a} = \lambda^{\alpha}h_{\alpha}{}^{a}$, and $\kappa^{\alpha} = -{2 \over \eta}\xi^{\alpha}_{a}(\Phi^{a} - Qs^{a})$. Likewise, eqn. \phihataonpsi\ implies that the action of ${\bf \Phi}^{\alpha}$ on $U^{(3)}$ can be written as
\eqnn \phihatalphaonpsi
$$
\eqalignno{
{\bf \Phi}^{\alpha}U^{(3)} &= \Phi^{\alpha} + Q\kappa^{\alpha} + {2 \over \eta}R_{\beta}{}^{\alpha}(\Phi^{\beta} + Q\kappa^{\beta}) & \phihatalphaonpsi
}
$$
where $\Phi^{\alpha} = \lambda^{\beta}h_{\beta}{}^{\alpha}$. This relation can be expressed in a more convenient way by using the alternative form for $R_{\alpha}{}^{\beta}$ 
\eqnn \alternativeralphabeta
$$
\eqalignno{
R_{\alpha}{}^{\beta} &= \bigg[{1 \over 12}(\lambda\gamma^{abcd})_{\alpha}(\lambda\gamma_{ab})^{\beta} + {2\over 3}(\lambda\gamma^{kd})_{\alpha}(\lambda\gamma^{c}{}_{k})^{\beta}  + {1\over 3}\lambda_{\alpha}(\lambda\gamma^{cd})^{\beta}  - {1\over 6}(\lambda\gamma^{cd})_{\alpha}\lambda^{\beta}\bigg](\bar{\lambda}\gamma_{cd}\bar{\lambda}) & \cr
& & \alternativeralphabeta
}
$$
or more compactly
\eqnn \rmorecompact
$$
\eqalignno{
R_{\alpha}{}^{\beta} &= {1\over 3}(\lambda\gamma_{ab})^{\beta}\rho_{ab\alpha} + {1\over 6}\lambda^{\beta}(\gamma^{cd}\lambda)_{\alpha}(\bar{\lambda}\gamma_{cd}\bar{\lambda}) & \rmorecompact
}
$$
where $\rho_{cd\alpha}$ was introduced in \rhotensor. Eqn. \phihatalphaonpsi\ then becomes
\eqnn \phihatalphaonu
$$
\eqalignno{
{\bf \Phi}^{\alpha}U^{(3)} &= \Phi^{\alpha} + Q\kappa^{\alpha} + (\lambda\gamma^{ab})^{\alpha}f_{ab} + \lambda^{\alpha}f & \phihatalphaonu
}
$$
where $f_{ab} = {2\over 3\eta}\rho_{ab\alpha}\tau^{\alpha} = -{1\over 6\eta}(\lambda\gamma_{abcd}\tau)(\bar{\lambda}\gamma^{cd}\bar{\lambda}) - {4\over 3\eta}(\lambda\gamma_{[bd}\tau)(\bar{\lambda}\gamma_{a]}{}^{d}\bar{\lambda}) - {2\over 3\eta}(\lambda\tau)(\bar{\lambda}\gamma_{ab}\bar{\lambda})$, $f={1\over 3\eta}(\lambda\gamma^{ab}\tau)(\bar{\lambda}\gamma_{ab}\bar{\lambda})$, and $\tau_{\alpha} = \Phi_{\alpha} + Q\kappa_{\alpha}$. This equation can be used to easily determine the behavior of ${\bf \Omega}_{ab}$ when acting on $U^{(3)}$ as
\eqnn \omegaabonu
$$
\eqalignno{
{\bf \Omega}_{cd}U^{(3)} &= \Omega_{cd} + 4Qf_{cd} + (\lambda\gamma_{cd}\lambda)t + (\lambda\gamma_{b[c}\lambda)t_{d]}{}^{b} + (\lambda\gamma_{b[c}\lambda)t^{b}{}_{d]} & \cr
& + {\cal R}_{cd}{}^{abmn}t_{abmn} - {16\over 3 \eta}(\lambda\gamma_{[ck}\lambda)(\bar{\lambda}\gamma_{d]}{}^{k}\bar{\lambda})Qf & \omegaabonu
}
$$
where $t = {1\over 3\eta}(\bar{\lambda}\gamma^{ab}\bar{\lambda})j_{ab}$, $t_{d}{}^{b} = -{8\over 3\eta}(\bar{\lambda}\gamma_{da}\bar{\lambda})j^{ab}$, $t_{abmn} = -{4 \over \eta}(\bar{\lambda}\gamma_{[ab}\bar{\lambda})j_{mn]}$, and $j_{ab} = \Omega_{ab} + 4Qf_{ab}$. As a consistency check, one can show that the right-hand side of eqn. \omegaabonu\ satisfies the relation $(\lambda\gamma^{cd})^{\alpha}Q({\bf \Omega}_{cd}U^{(3)}) = 0$. One can proceed in a similar way, and find the action of the other linear operators on $U^{(3)}$. Since we are ultimately interested in constructing the ghost number zero vertex operator, and as it will be seen later, these operators do not play any role for such a purpose, we will omit the details of these manipulations.

\medskip
 Now that we know the explicit action of the linear operators on $U^{(3)}$, we can determine the behavior of the non-linear operators when acting on $U^{(3)}$.

\medskip
\noindent
It is straightforward to show that eqns. \chatalpha, \chatalphaonpsi\ require that
\eqnn \chatalphabetaonpsi
$$
\eqalignno{
{\bf C}_{\alpha\beta}U^{(3)}  &=  C_{\beta\alpha} + (\lambda\gamma_{a})_{(\alpha}\tilde{t}_{\beta)}^{a} + {1\over 2}(\gamma^{a})_{\alpha\beta}\rho_{a} & \chatalphabetaonpsi
}
$$
where the tensor $\tilde{t}_{\beta}^{a}$ is given by
\eqnn \ttilde
$$
\eqalignno{
\tilde{t}_{\beta}^{a} &= {1\over \eta}(\lambda\gamma^{abc}C_{\beta})(\bar{\lambda}\gamma_{bc}\bar{\lambda}) + {1\over 2\eta}(\lambda\gamma^{apq}\gamma^{b})_{\beta}(\bar{\lambda}\gamma_{pq}\bar{\lambda})\rho_{b} + {3\over 2}{\bf C}_{\beta}\rho^{a} & \ttilde
}
$$
After some lengthy algebraic manipulations, it is possible to show that eqns. \chatalpha, \chata, \chatalphaonpsi\ and \chataonpsi\ imply that
\eqnn \chatalphabonpsi
$$
\eqalignno{
{\bf C}_{a \alpha}U^{(3)}
&= C_{a\alpha} - {1\over 2}D_{\alpha}\rho_{a} + Q\tilde{W}_{a\alpha} + (\lambda\gamma^{r})_{\alpha}\tilde{X}_{ra} + (\gamma_{ac}\lambda)_{\alpha}\tilde{Y}^{c} + (\lambda\gamma_{ac}\lambda)\tilde{Z}^{c}_{\alpha} & \chatalphabonpsi
}
$$
with the objects $\tilde{W}_{a\alpha}$, $\tilde{X}_{ra}$, $\tilde{Y}^{c}$, $\tilde{Z}_{\alpha}^{c}$ defined as
\eqnn \wtilde
\eqnn \xtilde
\eqnn \ytilde
\eqnn \ztilde
$$
\eqalignno{
\tilde{W}_{a\alpha} &= -{3\over 4}{\bf C}_{\alpha}\rho_{a} - {1\over 2\eta}(\lambda\gamma_{pqa}C_{\alpha})(\bar{\lambda}\gamma^{pq}\bar{\lambda}) + {1\over 4\eta}(\lambda\gamma_{pqa}\gamma^{b})_{\alpha}(\bar{\lambda}\gamma^{pq}\bar{\lambda})\rho_{b} & \wtilde\cr
\tilde{X}_{ra} &= {1\over 2\eta}(\lambda\gamma_{pqr}C_{a})(\bar{\lambda}\gamma^{pq}\bar{\lambda}) - {3\over 4}{\bf C}_{r}\rho_{a} + {1\over \eta}\xi_{a}^{\delta}C_{r\delta} - {1\over 2\eta}\xi_{a}^{\delta}D_{\delta}\rho_{r} - {3\over 2\eta}\xi_{a}^{\delta}Q{\bf C}_{\delta}\rho_{r}& \xtilde\cr
\tilde{Y}^{c} &= {1\over 2}s^{c} + {1\over \eta}(\lambda\gamma_{pr}\lambda)(\bar{\lambda}\gamma^{pc}\bar{\lambda})s^{r} & \ytilde\cr
\tilde{Z}_{\alpha}^{c} &= {3\over 4}{\bf C}_{\alpha}s^{c} - {1\over \eta}(\bar{\lambda}\gamma^{bc}\bar{\lambda})C_{b\alpha} + {1\over \eta}(\bar{\lambda}\gamma^{rc}\bar{\lambda})Q\bigg[{1\over \eta}(\lambda\gamma^{pqr}C_{\alpha})(\bar{\lambda}\gamma_{pq}\bar{\lambda}) + {1\over 2\eta}(\lambda\gamma^{pqr}\gamma^{b})_{\alpha}(\bar{\lambda}\gamma_{pq}\bar{\lambda})\rho_{b}\bigg] & \cr
& - {1\over 2\eta}(\gamma^{pq}\lambda)_{\alpha}(\bar{\lambda}\gamma_{pq}\bar{\lambda})s^{c} & \ztilde
}
$$
Equivalently, eqns. \chata, \chataonpsi\ yield
\eqnn \chatabonpsi
$$
\eqalignno{
{\bf C}_{ab}U^{(3)} &= C_{ab} + \partial_{[a}\rho_{b]} + Q\tilde{f}_{ab} + (\lambda\gamma_{[ah}\lambda)\tilde{p}^{h}{}_{b]} + (\lambda\gamma_{[bf}\lambda)q^{f}{}_{a]} + (\gamma_{ab}\lambda)_{\alpha}\tilde{v}^{\alpha}
& \chatabonpsi
}
$$
where the quantities $\tilde{f}_{rb}$, $\tilde{p}_{hb}$, $\tilde{q}_{fa}$, $\tilde{v}^{\alpha}$ take the form
\eqnn \ftilde
\eqnn \ptilde
\eqnn \qtilde
\eqnn \utilde
$$
\eqalignno{
\tilde{f}_{ab} &= {1\over \eta}(\lambda\gamma_{apq}C_{b})(\bar{\lambda}\gamma^{pq}\bar{\lambda}) - {3\over 2}{\bf C}_{a}\rho_{b} &\ftilde\cr
\tilde{p}_{hb} &= -{2\over \eta}(\bar{\lambda}\gamma^{s}{}_{h}\bar{\lambda})[C_{sb} + Q\tilde{f}_{sb} + {1\over 2}\partial_{s}\rho_{b}] & \ptilde \cr 
\tilde{q}_{fa} &= {1\over \eta}\xi_{a}^{\alpha}[-h_{\alpha f} + D_{\alpha}s_{f} + 3Q({\bf C}_{\alpha}s_{t})] & \qtilde \cr 
\tilde{v}^{\alpha} &= -{2\over \eta}\xi^{\alpha}_{f}(\phi^{f} - Qs^{f}) & \utilde
}
$$
See Appendix D for a more detailed discussion on the deduction of these formulae.

\medskip
As before, one could continue and compute the action of the other non-linear operators on $U^{(3)}$ by following the same line of reasoning developed above. Since these operators are irrelevant for our analysis in the next section, we will not discuss such computations.

\newsec Non-minimal Ghost Number Zero Vertex Operator

\seclab\secfour

In this section we will define the ghost number zero vertex operator by making use of the physical operators introduced in the previous section. This new operator will be shown to correctly satisfy a standard descent relation with $U^{(1)}$, i.e. $\{Q, U^{(0)}\} = \partial_{\tau}U^{(1)}$, as well as to reproduce the 2-particle superfield carrying ghost number three $U_{ij}^{(3)}$ defined in \tamingbghost, as a result of its action on a ghost number three single particle superfield.

\subsec Construction and consistency checks

\subseclab\subsecfourone

The ghost number zero vertex operator will be defined as follows
\eqnn \vertexoperatorgzero
$$
\eqalignno{
U^{(0)} &= 3\bigg[P^{a}{\bf C}^{b}\Omega_{ab} + (\lambda\gamma^{bc}\lambda)P^{a}h_{ab}{\bf \Phi}_{c} - (\lambda\gamma_{a})_{\delta}\lambda^{\alpha}T_{b\alpha}{}^{\delta}{\bf C}^{ab} + P^{a}{\bf C}_{b}\partial_{a}\phi^{b} - P^{a}{\bf C}_{\alpha}\partial_{a}\phi^{\alpha}\bigg]& \cr 
& &\vertexoperatorgzero
}
$$
where $[\Omega_{ab}, h_{ab}, T_{b\alpha}{}^{\delta}, \phi^{b}, \phi^{\alpha}]$ are the 11D supergravity superfields studied in section \sectwo, and $[{\bf C}_{b}, {\bf \Phi}_{c}, {\bf C}_{ab}, {\bf C}_{\alpha}]$ are the 11D physical operators discussed in section \secthree. Let us compute the action of the BRST charge on $U^{(0)}$. The first term in \vertexoperatorgzero\ results in
\eqnn \firsterm
$$
\eqalignno{
Q(P^{a}{\bf C}^{b}\Omega_{ab}) &= -P^{a}(\lambda\gamma^{bc}\lambda){\bf \Phi}_{c}\Omega_{ab} - {1\over 2}P^{a}{\bf C}^{b}R_{\alpha\beta,a}{}^{b}\lambda^{\alpha}\lambda^{\beta} & \cr
&= -P^{a}(\lambda\gamma^{bc}\lambda){\bf \Phi}_{c}\Omega_{ab} + P^{[a}{\bf C}^{b]}(\lambda\gamma_{b})_{\delta}T_{a\beta}{}^{\delta}\lambda^{\beta} & \firsterm
}
$$
Similarly, the second term transforms into
\eqnn \secondterm
$$
\eqalignno{
Q((\lambda\gamma^{bc}\lambda)P^{a}h_{ab}{\bf \Phi}_{c}) &= (\lambda\gamma^{bc}\lambda)P^{a}{\bf \Phi}_{c}(\Omega_{ab} + \partial_{a}\phi_{b}) & \secondterm
}
$$
The third term in turn contributes as
\eqnn \thirdterm
$$
\eqalignno{
Q((\lambda\gamma_{a})_{\delta}T_{b\alpha}{}^{\delta}\lambda^{\alpha}{\bf C}^{ab})&= (\lambda\gamma_{a})_{\delta}T_{b\alpha}{}^{\delta}\lambda^{\alpha}(-P^{[a}{\bf C}^{b]} + (\lambda\gamma^{[a c}\lambda){\bf h}_{c}{}^{b]} - (\lambda\gamma^{ab}{\bf \Phi})) & \cr
&= -(\lambda\gamma_{a})_{\delta}T_{b\alpha}{}^{\delta}\lambda^{\alpha}P^{[a}{\bf C}^{b]} & \thirdterm
}
$$
while the fourth term yields
\eqnn \fourthterm
$$
\eqalignno{
Q(P^{a}{\bf C}_{b}\partial_{a}\phi^{b}) &= {1\over 3}P^{a}P_{b}\partial_{a}\phi^{b} - P^{a}(\lambda\gamma_{bc}\lambda){\bf \Phi}^{c}\partial_{a}\phi^{b} - P^{a}{\bf C}_{b}\partial_{a}(\lambda\gamma_{b}\phi)& \fourthterm
}
$$
Finally, the last term produces
\eqnn \fifthterm
$$
\eqalignno{
Q(P^{a}{\bf C}_{\alpha}\partial_{a}\phi^{\alpha}) &= -{1\over 3}P^{a}d_{\alpha}\partial_{a}\phi^{\alpha} - P^{a}(\lambda\gamma^{b})_{\alpha}{\bf C}_{b}\partial_{a}\phi^{\alpha} + P^{a}{\bf C}_{\alpha}\partial_{a}\Omega^{\alpha}& \fifthterm
}
$$
Putting all together, one then learns that
\eqnn \descentrelation
$$
\eqalignno{
[Q, U^{(0)}] &= P^{a}\partial_{a}\bigg[P_{b}\phi^{b} + d_{\alpha}\phi^{\alpha} - w_{\alpha}\Omega^{\alpha}\bigg] & \cr
&= \partial_{\tau}U^{(1)} & \descentrelation
}
$$
as stated.

\subsec The two-particle superfield

\subseclab\subsecfourtwo

 In \tamingbghost, the simplification of the b-ghost was used for computing the 2-particle superfield whose BRST variation reproduces the current: $(\lambda\gamma^{ab}\lambda)\phi_{i,a}\phi_{j,b}$, with $(i,j)$ particle labels. Next, we will reproduce such a multi-particle superfield up to BRST-exact and contact terms, from the operator action of the ghost number zero vertex displayed in eqn. \vertexoperatorgzero\ associated to the particle $i$, $U^{(0)}_{i}$, on the ghost number three vertex operator associated to the particle $j$, $U^{(3)}_{j}$. 

\medskip
Let us compute $U^{(0)}_{i}(U^{(3)}_{j})$: The first term in \vertexoperatorgzero\ contributes as
\eqnn \firsttermmulti
$$
\eqalignno{
P^{a}{\bf C}^{b}\Omega_{i,ab}(U^{(3)}_{j}) &= -\Omega_{i,ab}(\partial^{a}C^{b}_{j}+(\lambda\gamma^{bk}\lambda)\partial^{a}s_{j,k} - Q\partial^{a}\rho_{j}^{b})& \cr
&=  -\Omega_{i,ab}(\partial^{a}C^{b}_{j}+(\lambda\gamma^{bk}\lambda)\partial^{a}s_{j,k}) + Q(-\Omega_{i,ab}\partial^{a}\rho_{j}^{b}) + (\lambda\gamma_{a})_{\delta}\lambda^{\alpha}T_{i,b \alpha}{}^{\delta}\partial^{a}\rho_{j}^{b}&\cr
& &\firsttermmulti
}
$$
where we made use of eqn. \chataonpsi\ and the equation of motion: $R_{(\alpha\beta),b}{}^{c} + 2(\gamma^{c})_{\gamma(\beta}T_{|b| \alpha)}{}^{\gamma} = 0$, see \refs{\maxthesis, \elevendsuperspaceexpansion} for details. In turn, eqns. \eomfour\ and \phihataonpsi\ imply the following for the second term of \vertexoperatorgzero
\eqnn \secondtermmulti
$$
\eqalignno{
(\lambda\gamma^{bc}\lambda)h_{i,ab}P^{a}{\bf \Phi}_{c}(U^{(3)}_{j})&= (\lambda\gamma^{bc}\lambda)h_{i,ab}[\partial^{a}\phi_{j,c} - Q\partial^{a}s_{j,c}]  & \cr
&= (\lambda\gamma^{bc}\lambda)h_{i,ab}\partial^{a}\phi_{j,c} + Q[-(\lambda\gamma^{bc}\lambda)h_{i,ab}\partial^{a}s_{j,c}]& \cr
& + (\lambda\gamma^{bc}\lambda)\partial_{a}\phi_{i,b}\partial^{a}s_{j,c}  + (\lambda\gamma^{bc}\lambda)\Omega_{i,ab}\partial^{a}s_{j,c} & \secondtermmulti
}
$$
\noindent The third term of the operator \vertexoperatorgzero\ results in
\eqnn \thirdtermmulti
$$
\eqalignno{
-(\lambda\gamma^{a})_{\delta}\lambda^{\alpha}T_{i,b \alpha}{}^{\delta}{\bf C}^{ab}(U^{(3)}_{j})&= -(\lambda\gamma_{a})_{\delta}\lambda^{\alpha}T_{i,b\alpha}{}^{\delta}C_{j}^{ab} - (\lambda\gamma_{a})_{\delta}\lambda^{\alpha}T_{i,b\alpha}{}^{\delta}\partial^{a}\rho^{b}_{j} - (\lambda\gamma_{a})_{\delta}\lambda^{\alpha}T_{i,b\alpha}{}^{\delta}Q\tilde{f}^{ab}_{j}  & \cr
&= -(\lambda\gamma_{a})_{\delta}\lambda^{\alpha}T_{i,b\alpha}{}^{\delta}C_{j}^{ab} - (\lambda\gamma_{a})_{\delta}\lambda^{\alpha}T_{i,b\alpha}{}^{\delta}\partial^{a}\rho^{b}_{j} + Q[- (\lambda\gamma_{a})_{\delta}\lambda^{\alpha}T_{i,b\alpha}{}^{\delta}\tilde{f}^{ab}_{j})]& \cr
& &\thirdtermmulti
}
$$
 whereas the last two terms contribute as
\eqnn \fourthtermmulti
\eqnn \fifthtermmulti
$$
\eqalignno{
-P^{a}\partial_{a}\phi_{i,b}{\bf C}^{b}(U^{(3)}_{j}) &=
-\partial_{a}\phi_{i,b}[\partial^{a}C_{j}^{b} + (\lambda\gamma^{bk}\lambda)\partial^{a}s_{j,k} - Q\partial^{a}\rho^{b}_{j}] & \cr
&= -\partial_{a}\phi_{i,b}[\partial^{a}C_{j}^{b} + (\lambda\gamma^{bk}\lambda)\partial^{a}s_{j,k}] + Q[-\partial_{a}\phi_{i,b}\partial^{a}\rho_{j}^{b}] + \partial_{a}\phi_{i}^{\delta}(\lambda\gamma^{b})_{\delta}\partial^{a}\rho_{j}^{b}& \cr
& &\fourthtermmulti \cr
-P^{a}\partial_{a}\phi_{i}^{\alpha}{\bf C}_{\alpha}(U^{(3)}_{j}) &= -\partial_{a}\phi_{i}^{\alpha}[\partial^{a}C_{j,\alpha} + (\lambda\gamma^{r})_{\alpha}\partial^{a}\rho_{j,r}] & \fifthtermmulti
}
$$
where we used eqn. \eomone\ in the second line of eqn. \fourthtermmulti. Collecting all terms, one then finds that
\eqnn \ghostzeroonghostthree
$$
\eqalignno{
U^{(0)}_{i}(U^{(3)}_{j}) &= 3\bigg[-\Omega_{i,ab}\partial^{a}C_{j}^{b} + (\lambda\gamma^{bc}\lambda)h_{i,ab}\partial^{a}\phi_{j,c} - (\lambda\gamma_{a})_{\delta}\lambda^{\alpha}T_{i,b\alpha}{}^{\delta}C_{j}^{ab} - \partial_{a}\phi_{i,b}\partial^{a}C_{j}^{b} &\cr
& - \partial_{a}\phi_{i}^{\alpha}\partial^{a}C_{j,\alpha}\bigg] + Q\bigg[3(-\Omega_{i,ab}\partial^{a}\rho_{j}^{b} - (\lambda\gamma^{bc}\lambda)h_{i,ab}\partial^{a}s_{j,c} - \partial_{a}\phi_{i,b}\partial^{a}\rho_{j}^{b} &\cr
& - (\lambda\gamma_{a})_{\delta}\lambda^{\alpha}T_{i,b\alpha}{}^{\delta}\tilde{f}_{j}^{ab})\bigg] & \ghostzeroonghostthree
}
$$
which after symmetrization in particle labels coincides with the 2-particle vertex obtained in \tamingbghost\ from the action of the b-ghost, up to BRST-exact and contact terms.

\newsec Discussions

\seclab \secsix

 In this paper we have constructed new physical operators consistent with the equations of motion of 11D supergravity, and we have introduced a ghost number zero vertex operator which presents a strikingly simple form when expressed in terms of the former. This operator along with the ghost number one vertex were successfully shown to satisfy a standard descent relation, and it was also verified that the ghost number zero operator reproduces the 2-particle ghost number three pure spinor superfield discussed in \tamingbghost, when acting on a single particle ghost number three vertex operator.

\medskip
It is interesting to see that some non-linear physical operators of section 3 exhibit a simple structure, and some do not. For instance, ${\bf C}_{\alpha\beta}$ in the first line of eqn. \quadraticop\ can be written as a single composition of the linear operators ${\bf C}_{\alpha}$ and ${\bf C}_{\beta}$, while ${\bf h}_{\alpha}{}^{\delta}$ in the fourth line of eqn. \quadraticop\ presents extra terms that cannot be written as a simple composition of linear physical operators. This might be due to normal ordering issues since the linear physical operators do not commute with each other, and the implicit prescription used in this work, which was adopted based on the most basic definition that one could assign to the physical operators related to the super-3-form component of lowest mass dimension, namely [${\bf C}_{\alpha}$, ${\bf C}_{\alpha\beta}$, ${\bf C}_{\alpha\beta\delta}$]. Although physical operators might not be unique in this regard, it is important to emphasize that the algebraic forms found for these in this paper are consistent and compatible with the defining relations of section 3. We intend to investigate other possible normal ordering schemes and related problems in future work.

\medskip
 Our results open up numerous avenues for future research. In particular, the ghost number zero vertex operator of eqn. \vertexoperatorgzero\ can be used for testing the 11D pure spinor wordline correlator proposed in \Grassi. This is done in our companion paper which explicitly computes, for the first time, the four-point function in 11D pure spinor superspace \amplitudeeleven. Furthermore, following the ideas developed in \refs{\supertwistorambitwistor,\guillengarciaparticle,\guillengarciaambitwistor,\guillenchiral} which introduce an ambitwistor string from a worldline model, it would be interesting to explore a possible extension of our 11D worldline construction to the ambitwistor framework. Likewise, it is intriguing to see that the particle-limit of the supermembrane ghost number zero vertex operator does not reduce to the vertex operator presented in this work. This easily follows from the fact that the supermembrane operator discussed in \pssupermembrane\ only contains minimal variables in its definition. This might have strong implications in the construction of the pure spinor supermembrane framework. We leave this and related questions for future investigation.

\bigskip \noindent{\bf Acknowledgements:} MG is grateful to Lionel Mason and Nathan Berkovits for collaboration on related topics. MG would also like to thank Martin Cederwall, Henrik Johansson, Renann Jusinskas, Carlos Mafra and Oliver Schlotterer for enlightening discussions on 11D supergravity and pure spinors. EV wants to thank Kostas Rigatos, Thiago Fleury, Rennan Jusinskas and Humberto Gomez for useful discussions on this topic. We also thank ICTP-SAIFR for their hospitality during the workshop on Modern Amplitude Methods for Gauge and Gravity Theories, and Nordita for hosting Eurostrings 2025 where this work was completed. The work of MG was partially funded by the European Research Council under ERC-STG-804286 UNISCAMP, and by the Knut and Alice Wallenberg Foundation under the grant KAW 2018.0162 (Exploring a Web of Gravitational Theories through Gauge-Theory Methods) and the Wallenberg AI, Autonomous Systems and Software Program (WASP). The work of EV was supported in part by ICTP-SAIFR FAPESP grant 2019/21281-4 and by FAPESP grant 2022/00940-2.


\appendix{A}{Commutation Relations of Physical Operators}

\seclab\appendixa

Unlike in 10D, the 11D physical operators obey intricate (anti)commutation relations. To see this, let us compute some commutators involving the operator ${\bf C}_{\alpha}$. One can start with the self-commutator $[{\bf C}_{\alpha}, {\bf C}_{\beta}]$:
\eqnn \commutatorcalphacbeta
$$
\eqalignno{
[{\bf C}_{\alpha}, {\bf C}_{\beta}] &= {1\over 9}\left(K_{\alpha}{}^{\mu}[w_{\mu}, K_{\beta}{}^{\delta}]w_{\delta} - (\alpha \leftrightarrow \beta)\right) & \commutatorcalphacbeta
}
$$
The following the relation
\eqnn \relationone
$$
\eqalignno{
[w_{\mu}, K_{\beta}{}^{\delta}] &= {4\over \eta^{2}}(\gamma^{mn}\lambda)_{\mu}(\bar{\lambda}\gamma_{mn}\bar{\lambda})\xi_{a}^{\delta}(\lambda\gamma^{a})_{\beta} + {1\over \eta}(\gamma^{abc})^{\delta}{}_{\mu}(\bar{\lambda}\gamma_{bc}\bar{\lambda})(\lambda\gamma_{a})_{\beta} - {2\over \eta}\xi_{a}^{\delta}(\gamma^{a})_{\beta\mu} & \cr
& & \relationone
}
$$
then enables one to conclude that
\eqnn \commutatorcalphacbetafinal
$$
\eqalignno{
[{\bf C}_{\alpha}, {\bf C}_{\beta}] &= {2\over 3\eta}(\bar{\lambda}\gamma^{ab}\bar{\lambda})(\lambda\gamma^{c})_{[\alpha}(\gamma_{abc})_{\beta]}{}^{\delta}{\bf C}_{\delta} + {4\over 9\eta^{2}}(\lambda\gamma^{cd}w)(\bar{\lambda}\gamma_{cd}\bar{\lambda})(\bar{\lambda}\gamma^{fa}\bar{\lambda})(\lambda\gamma_{f})_{\alpha}(\lambda\gamma_{a})_{\beta} & \cr
& & \commutatorcalphacbetafinal
}
$$

\medskip
Similarly, one can calculate the commutator $[{\bf C}_{\alpha}, {\bf C}_{a}]$:
\eqnn \commchatalphachata
$$
\eqalignno{
[{\bf C}_{\alpha}, {\bf C}_{a}] &= -{2\over 3\eta}(\gamma^{pq}\lambda)_{\alpha}(\bar{\lambda}\gamma_{pq}\bar{\lambda}){\bf C}_{a} - {2 \over \eta}\bigg[-{1\over 6}(\gamma_{apq})^{\delta}{}_{\alpha}(\bar{\lambda}\gamma_{pq}\bar{\lambda})& \cr
& + {1\over 3\eta}(\lambda\gamma^{cq})^{\delta}(\bar{\lambda}\gamma_{cq}\bar{\lambda})(\bar{\lambda}\gamma^{ar}\bar{\lambda})(\lambda\gamma_{r})_{\alpha}\bigg]d_{\delta} + \ldots & \commchatalphachata
}
$$
where we used eqn. \chata, and the fact that
\eqnn \identityone
\eqnn \identitytwo
$$
\eqalignno{
[{\bf C}_{\alpha}, \xi_{a}{}^{\beta}] &= -{1\over 6}(\gamma_{apq})^{\delta}{}_{\alpha}(\bar{\lambda}\gamma_{pq}\bar{\lambda}) + {1\over 3\eta}(\lambda\gamma^{cq})^{\delta}(\bar{\lambda}\gamma_{cq}\bar{\lambda})(\bar{\lambda}\gamma^{ar}\bar{\lambda})(\lambda\gamma_{r})_{\alpha} & \identityone \cr
[{\bf C}_{\alpha}, \eta] &= {2\over 3}(\gamma^{ab}\lambda)_{\alpha}(\bar{\lambda}\gamma_{ab}\bar{\lambda})
& \identitytwo
}
$$
Ellipsis in eqn. \commchatalphachata\ stands for the contributions coming from the piece proportional to ${\bf C}_{\alpha}$ in ${\bf C}_{a}$. These terms will be identically cancelled after realizing that the $d_{\alpha}$ variables present in \commchatalphachata\ are the same as those appearing in the definition of ${\bf C}_{a}$, and thus one can use the relation\foot{Similar arguments have been used in \tamingbghost, where the 11D b-ghost was expressed in terms of a fermionic vector in a quite compact way.}
\eqnn \dalphaintermsofchata
$$
\eqalignno{
d_{\alpha}  + 3[Q, {\bf C}_{\alpha}] &= -3(\gamma^{a}\lambda)_{\alpha}{\bf C}_{a} & \dalphaintermsofchata
}
$$
In this manner, one finds
\eqnn \commchatalphachatapartii
$$
\eqalignno{
[{\bf C}_{\alpha}, {\bf C}_{a}] &= -{2\over 3 \eta}(\gamma^{pq}\lambda)_{\alpha}(\bar{\lambda}\gamma_{pq}\bar{\lambda}){\bf C}_{a} + {1\over 3\eta}(\gamma_{apq}\gamma^{s}\lambda)_{\alpha}(\bar{\lambda}\gamma^{pq}\bar{\lambda}){\bf C}_{s} - {2\over \eta}\xi_{a}^{\beta}[{\bf  C}_{\alpha},[Q,{\bf C}_{\beta}]] & \cr
&= -{1\over 3\eta}(\gamma^{s}\gamma_{apq}\lambda)_{\alpha}(\bar{\lambda}\gamma^{pq}\bar{\lambda}){\bf C}_{s} - {2\over \eta}\xi_{a}^{\beta}[{\bf  C}_{\alpha},[Q,{\bf C}_{\beta}]] &
\commchatalphachatapartii
}
$$

\appendix{B}{Quadratic Physical Operators}

\seclab\appendixb

This Appendix shows that the identifications made in eqn. \quadraticop\ are compatible with the defining relations \morephysophalphab
-\morephysopcab. 

\medskip
  One starts with the definition ${\bf C}_{\alpha\beta} = {3\over 2}{\bf C}_{(\alpha} \circ {\bf C}_{\beta)}$, and plug it into eqn. \morephysopcalphabeta\ to find that
\eqnn \checkingcalphabeta
$$
\eqalignno{
{3\over 2}[Q, {\bf C}_{(\alpha}]{\bf C}_{\beta)} + {3\over 2}{\bf C}_{(\alpha}[Q, {\bf C}_{\beta)}] &= -D_{(\alpha}{\bf C}_{\beta)} - 2(\lambda\gamma^{a})_{(\alpha}{\bf C}_{a\beta)} - {1\over 2}(\gamma^{a})_{\alpha\beta}{\bf C}_{a} &\checkingcalphabeta
}
$$ 
Eqn. \drchatalpha\ then implies that \checkingcalphabeta\ becomes
\eqnn \checkingcalphabetapartii
$$
\eqalignno{
-{3\over 2}(\lambda\gamma^{a})_{(\alpha}{\bf C}_{a}\circ {\bf C}_{\beta)} - {3\over 2}{\bf C}_{(\alpha}\circ (\lambda\gamma^{a})_{\beta)}{\bf C}_{a} &= -2(\lambda\gamma^{a})_{(\alpha}{\bf C}_{a\beta)} - {1\over 2}(\gamma^{a})_{\alpha\beta}{\bf C}_{a} & \checkingcalphabetapartii
}
$$
It is not hard to check that
\eqnn \identitythree
$$
\eqalignno{
[{\bf C}_{\alpha}, (\lambda\gamma^{a})_{\beta}] &= {1\over 3}(\gamma^{a})_{\alpha\beta} + {1\over 3\eta}(\lambda\gamma^{pqr}\gamma^{a})_{\beta}
(\bar{\lambda}\gamma_{pq}\bar{\lambda})(\lambda\gamma_{r})_{\alpha} & \identitythree
}
$$
Therefore, eqn. \checkingcalphabetapartii\ simplifies to
\eqnn \checkingcalphabetapartiii
$$
\eqalignno{
-{3\over 2}(\lambda\gamma^{a})_{(\alpha}{\bf C}_{a}\circ {\bf C}_{\beta)} -{3\over 2}(\lambda\gamma^{a})_{(\alpha}{\bf C}_{\beta)} \circ {\bf C}_{a} - {1\over 2\eta}(\lambda\gamma^{pqr}\gamma^{a})_{(\beta}(\bar{\lambda}\gamma_{pq}\bar{\lambda})(\lambda\gamma_{r})_{\alpha)}{\bf C}_{a} &= -2(\lambda\gamma^{a})_{(\alpha} {\bf C}_{a\beta)} & \cr
& & \checkingcalphabetapartiii
}
$$
This equation is solved for
\eqnn \checkingcalphabetapartiv
$$
\eqalignno{
{\bf C}_{a\alpha} &= {3\over 4}{\bf C}_{a}\circ {\bf C}_{\alpha} + {3\over 4}{\bf C}_{\alpha}\circ {\bf C}_{a} + {1\over 4\eta}(\lambda\gamma^{apq}\gamma^{b})_{\alpha}(\bar{\lambda}\gamma^{pq}\bar{\lambda}){\bf C}_{b} & \checkingcalphabetapartiv
}
$$

\medskip
 One can now use eqn. \checkingcalphabetapartiv\ to write eqn. \morephysopcalphab\ as
\eqnn \checkingcalphab
$$
\eqalignno{
-{3\over 4}[Q, {\bf C}_{\delta}] {\bf C}_{a} - {3\over 4}&{\bf C}_{\delta}\{Q, {\bf C}_{a}\} -{3\over 4}\{Q, {\bf C}_{a}\} {\bf C}_{\delta} + {3\over 4}{\bf C}_{a}[Q, {\bf C}_{\delta}] + \bigg\{Q,{1\over 2\eta}\xi^{a\delta}(\gamma^{b})_{\delta\alpha}{\bf C}_{b}\bigg\} & \cr
&= {1\over 2}(D_{\delta}{\bf C}_{a} - \partial_{a}{\bf C}_{\delta}) 
 - (\lambda\gamma^{b})_{\delta}{\bf C}_{ba} + {1\over 2}(\lambda\gamma_{ab}\lambda){\bf h}_{\delta}{}^{b} - (\lambda\gamma_{ab})_{\delta}{\bf \Phi}^{b}  &\checkingcalphab
}
$$
which can easily be seen to take the form
\eqnn \checkingcalphabpartii
$$
\eqalignno{
{3 \over 4}&(\lambda\gamma^{b})_{\delta}\bigg[{\bf C}_{b}\circ {\bf C}_{a} - {\bf C}_{a}\circ {\bf C}_{b}\bigg]  + {3\over 4}(\lambda\gamma_{ab}\lambda)\bigg[{\bf \Phi}^{b} \circ {\bf C}_{\delta} + {\bf C}_{\delta}\circ {\bf \Phi}^{b}\bigg] + {3 \over 4}[{\bf C}_{\delta}, (\lambda\gamma_{ab}\lambda)]\circ {\bf \Phi}^{b} - {1\over 4}\{{\bf C}_{a}, D_{\delta}\} &\cr
&- {3\over 4}[{\bf C}_{a}, (\lambda\gamma^{b})_{\delta}]\circ {\bf C}_{b} + \bigg\{Q,{1\over 2\eta}\xi^{a\alpha}(\gamma^{b})_{\alpha\delta}{\bf C}_{b}\bigg\}  = 
-(\lambda\gamma^{b})_{\delta}{\bf C}_{ba} + {1\over 2}(\lambda\gamma_{ab}\lambda) {\bf h}_{\delta}{}^{b} - (\lambda\gamma_{ab})_{\delta}{\bf \Phi}^{b}&\cr
& & \checkingcalphabpartii
}
$$
The use of the identities
\eqnn \identitiesfourzero
\eqnn \identitiesfoura
\eqnn \identitiesfourb
$$
\eqalignno{
\{{\bf C}_{a}, D_{\delta}\} &= {2\over 3\eta}\xi_{a}{}^{\alpha}(\gamma^{b})_{\alpha\delta}\partial_{b} & \identitiesfourzero\cr
[{\bf C}_{\delta}, (\lambda\gamma_{ab}\lambda)] &= {2\over 3}(\gamma_{ab}\lambda)_{\delta} + {2\over 3 \eta}(\lambda\gamma_{abpqr}\lambda)(\bar{\lambda}\gamma^{pq}\bar{\lambda})(\lambda\gamma^{r})_{\delta} & \identitiesfoura\cr
[{\bf C}_{a}, (\lambda\gamma^{b})_{\delta}] &= {2\over 3}[Q, {1\over \eta}\xi_{a}^{\alpha}(\gamma^{b})_{\alpha\delta}] - {4\over 3\eta} (\lambda\gamma_{at}\lambda)(\bar{\lambda}\gamma^{pt}\bar{\lambda})[Q, {1\over \eta}\xi_{p}^{\alpha}(\gamma^{b})_{\alpha\delta}]& \identitiesfourb
}
$$
along with eqn. \phihata, imply that eqn. \checkingcalphabpartii\ simplifies to
\eqnn \checkingcalphabpartiiiintermediate
$$
\eqalignno{
{3 \over 4}&(\lambda\gamma^{b})_{\delta}\bigg[{\bf C}_{b}\circ {\bf C}_{a} - {\bf C}_{a}\circ {\bf C}_{b}\bigg]  + {3\over 4}(\lambda\gamma_{ab}\lambda)\bigg[{\bf \Phi}^{b} \circ {\bf C}_{\delta} + {\bf C}_{\delta}\circ {\bf \Phi}^{b}\bigg] - {1\over 2}(\gamma_{ab}\lambda)_{\delta}{\bf \Phi}^{b} & \cr
&- {1\over 4\eta}(\gamma^{b}\gamma_{apq}\lambda)_{\delta}(\bar{\lambda}\gamma^{pq}\bar{\lambda})(\lambda\gamma_{bc}\lambda){\bf \Phi}^{c} - {2\over \eta^{2}}(\lambda\gamma_{at}\lambda)(\bar{\lambda}\gamma^{pt}r)\xi_{p}^{\alpha}(\gamma^{b})_{\alpha\delta}{\bf C}_{b} = 
-(\lambda\gamma^{b})_{\delta}{\bf C}_{ba} + {1\over 2}(\lambda\gamma_{ab}\lambda) {\bf h}_{\delta}{}^{b}  & \cr
& & \checkingcalphabpartiiiintermediate 
}
$$
where the last term in the first line of \checkingcalphabpartii\ as well as the BRST-exact term, were exactly cancelled by the contributions of the first term in \identitiesfourb. Hence, one finds that
\eqnn \checkingcalphabpartiii
$$
\eqalignno{
{3 \over 4}(\lambda\gamma^{b})_{\delta}[{\bf C}_{b}\circ {\bf C}_{a} - {\bf C}_{a}\circ {\bf C}_{b}]  + {3 \over 4}(\lambda\gamma_{ab}\lambda)\bigg[&{\bf C}_{\delta}\circ {\bf \Phi}^{b}+ {\bf \Phi}^{b} \circ {\bf C}_{\delta} - {2\over 3 \eta}(\gamma^{pq}\lambda)_{\delta}(\bar{
\lambda}\gamma_{pq}\bar{\lambda}){\bf \Phi}^{b} & \cr - {8\over 3\eta^{2}}(\bar{\lambda}\gamma^{pb}r)\xi_{p}^{\epsilon}(\gamma^{c})_{\epsilon\delta}{\bf C}_{c}\bigg]  
& = 
-(\lambda\gamma^{b})_{\delta}{\bf C}_{ba} + {1\over 2}(\lambda\gamma_{ab}\lambda) {\bf h}_{\delta}{}^{b} & \checkingcalphabpartiii}
$$
which is automatically solved by
\eqnn \soluone
\eqnn\solutwo
$$
\eqalignno{
{\bf C}_{ba} &= -{3\over 4}{\bf C}_{b}\circ {\bf C}_{a} + {3\over 4}{\bf C}_{a}\circ {\bf C}_{b} & \soluone\cr
{\bf h}_{\delta}{}^{b} &= {3\over 2}{\bf C}_{\delta}\circ {\bf \Phi}^{b} + {3\over 2}{\bf \Phi}^{b}\circ {\bf C}_{\delta} - {1\over \eta}(\gamma^{pq}\lambda)_{\delta}(\bar{\lambda}\gamma_{pq}\bar{\lambda}){\bf \Phi}^{b} - {4\over \eta^{2}}(\bar{\lambda}\gamma^{pb}r)\xi_{p}^{\epsilon}(\gamma^{c})_{\epsilon\delta}{\bf C}_{c} & \solutwo
}
$$

\medskip 
In a similar manner, one can subtitute eqn. \soluone\ into eqn. \morephysopcab\ as
\eqnn \checkingcalcb
$$
\eqalignno{
-{3\over 4}\{Q, {\bf C}_{a}\}{\bf C}_{b} + {3\over 4}{\bf C}_{a}\{Q, {\bf C}_{b}\} + {3\over 4}\{Q, {\bf C}_{b}\}{\bf C}_{a} - {3\over 4}{\bf C}_{b}\{Q, {\bf C}_{a}\} &= -\partial_{[a}{\bf C}_{b]} + (\lambda\gamma_{[ac}\lambda){\bf h}_{b]}{}^{c} - (\lambda\gamma_{ab})_{\delta}{\bf \Phi}^{\delta} &\cr
& & \checkingcalcb
}
$$
to learn that
\eqnn \checkingcalcbpartii
$$
\eqalignno{
{3\over 4}(\lambda\gamma_{ac}\lambda){\bf \Phi}^{c}\circ {\bf C}_{b} - {3\over 4}{\bf C}_{a}(\lambda\gamma_{bc}\lambda)\circ {\bf \Phi}^{c} - {3\over 4}(\lambda\gamma_{bc}\lambda){\bf \Phi}^{c}\circ {\bf C}_{a} + {3\over 4}{\bf C}_{b}(\lambda\gamma_{ac}\lambda)\circ {\bf \Phi}^{c} &= (\lambda\gamma_{[ac}\lambda){\bf h}_{b]}{}^{c} - (\lambda\gamma_{ab})_{\delta}{\bf \Phi}^{\delta} & \cr
& & \checkingcalcbpartii
}
$$
The use of the relation
\eqnn \identityfive
$$
\eqalignno{
[{\bf C}_{a}, (\lambda\gamma^{bd}\lambda)] &= -{2\over 3\eta}(\lambda\gamma_{k}\gamma_{a}{}^{pq}\lambda)(\bar{\lambda}\gamma_{pq}\bar{\lambda})(\lambda\gamma^{bdijk}\lambda)[Q, {1 \over \eta}(\bar{\lambda}\gamma_{ij}\bar{\lambda})] & \identityfive
}
$$
which in turn implies that
\eqnn \identityfivepartii
$$
\eqalignno{
[{\bf C}_{a}, (\lambda\gamma^{bd}\lambda)]{\bf \Phi}_{d} &={2\over 3}(\lambda\gamma_{a}{}^{b})_{\alpha}{\bf \Phi}^{\alpha} & \identityfivepartii
}
$$
allows one to simplify eqn. \checkingcalcbpartii\ to the form
\eqnn \checkingcalcbpartiii
$$
\eqalignno{
{3\over 2}(\lambda\gamma_{[ac}\lambda)\bigg[{\bf \Phi}^{c}\circ {\bf C}_{b]} + {\bf C}_{b]}\circ {\bf \Phi}^{c} \bigg]&= (\lambda\gamma_{[ac}\lambda){\bf h}_{b]}{}^{c} & \checkingcalcbpartiii
}
$$
where the last term of eqn. \checkingcalcbpartii\ was identically cancelled by the double contribution of the commutator \identityfivepartii\ appearing on the left-hand side of eqn. \checkingcalcbpartii. Thus, one concludes that
\eqnn \soluthree
$$
\eqalignno{
{\bf h}_{b}{}^{c} &= {3\over 2}({\bf \Phi}^{c}\circ {\bf C}_{b} + {\bf C}_{b}\circ {\bf \Phi}^{c}) & \soluthree
}
$$

\medskip
Analogously, one can use eqn. \solutwo\ to write eqn. \morephysophalphab\ in the form
\eqnn \firsteqnap
$$
\eqalignno{
{3\over 2}&[Q, {\bf C}_{\alpha}]{\bf \Phi}^{a} + {3\over 2}{\bf C}_{\alpha}[Q, {\bf \Phi}^{a}] + {3\over 2}[Q, {\bf \Phi}^{a}]{\bf C}_{\alpha} + {3\over 2}{\bf \Phi}^{a}[Q, {\bf C}_{\alpha}] + [Q, -{1\over \eta}(\gamma^{pq}\lambda)_{\alpha}(\bar{\lambda}\gamma_{pq}\bar{\lambda}){\bf \Phi}^{a}] &\cr
&+ [Q, -{4\over \eta^{2}}(\bar{\lambda}\gamma^{pa}r)\xi_{p}^{\epsilon}(\gamma^{c})_{\epsilon\alpha}{\bf C}_{c}] = -D_{\alpha}{\bf \Phi}^{a} + (\lambda\gamma^{a})_{\delta}{\bf h}_{\alpha}{}^{\delta} + (\gamma^{a}{\bf \Phi})_{\alpha} - (\lambda\gamma^{b})_{\alpha}{\bf h}_{b}{}^{a} & \firsteqnap
}
$$
After making use of eqns. \drchatalpha, \drphihata\ and \soluthree, one learns that eqn. \firsteqnap\ can be cast as
\eqnn \firsteqnapcontinue
$$
\eqalignno{
-{1\over 2}[&{\bf \Phi}^{a},D_{\alpha}] - {3\over 2}[{\bf \Phi}^{a},(\lambda\gamma^{c})_{\alpha}]\circ{\bf C}_{c} + {1\over 2}(\gamma^{a}{\bf \Phi})_{\alpha} + {3\over 2}(\lambda\gamma^{a})_{\delta}\bigg[{\bf C}_{\alpha}\circ {\bf \Phi}^{\delta} + {\bf \Phi}^{\delta}\circ {\bf C}_{\alpha}\bigg] & \cr
&+ [Q, -{1\over \eta}(\gamma^{pq}\lambda)_{\alpha}(\bar{\lambda}\gamma_{pq}\bar{\lambda}){\bf \Phi}^{a}] + [Q, -{4\over \eta^{2}}(\bar{\lambda}\gamma^{pa}r)\xi_{p}^{\epsilon}(\gamma^{c})_{\epsilon\alpha}{\bf C}_{c}] =   (\lambda\gamma^{a})_{\delta}{\bf h}_{\alpha}{}^{\delta} + (\gamma^{a}{\bf \Phi})_{\alpha}& \cr
& & \firsteqnapcontinue
}
$$
where we used that
\eqnn \firsteqnapcontinuetwo
$$
\eqalignno{
[{\bf C}_{\alpha}, (\lambda\gamma^{a})_{\delta}]{\bf \Phi}^{\delta} &= {1\over 3}(\gamma^{a}{\bf \Phi})_{\alpha} & \firsteqnapcontinuetwo
}
$$
as a consequence of ${\bf \Phi}^{\delta}$ being proportional to $\xi_{a}^{\delta}$. It is easy to check that
\eqnn \identityfirst
\eqnn \identitysecond
$$
\eqalignno{
[{\bf \Phi}^{a}, D_{\alpha}] &= -{8\over 3\eta^{2}}(\bar{\lambda}\gamma^{ac}r)\xi_{c}^{\delta}(\gamma^{s})_{\delta\alpha}\partial_{s} & \identityfirst\cr
[{\bf \Phi}^{a}, (\lambda\gamma^{c})_{\alpha}] &= -{8\over 3\eta}(\bar{\lambda}\gamma^{ad}r)[Q, {1\over \eta}\xi_{d}^{\delta}(\gamma^{c})_{\delta\alpha}] - {32\over 3\eta^{3}}(\bar{\lambda}\gamma^{ad}r)(\lambda\gamma_{dt}\lambda)(\bar{\lambda}\gamma^{pt}r)\xi_{p}^{\delta}(\gamma^{c})_{\delta\alpha}& \cr
& & \identitysecond
}
$$
In order to see how the BRST-exact terms in \firsteqnapcontinue\ exactly cancel out, one needs to carefully manipulate the terms on the right-hand side of eqn. \identitysecond. Indeed, the first term can be written in the more convenient way
\eqnn \identitythird
$$
\eqalignno{
-{8\over 3\eta}(\bar{\lambda}\gamma^{ad}r)[Q, {1\over \eta}\xi_{d}^{\delta}(\gamma^{s})_{\delta\alpha}]{\bf C}_{s} &= [Q, {8 \over 3\eta^{2}}(\bar{\lambda}\gamma^{ad}r)\xi_{d}^{\delta}(\gamma^{s})_{\delta\alpha}{\bf C}_{s}] + {16\over 3\eta^{3}}\phi(\bar{\lambda}\gamma^{ad}r)\xi_{d}^{\delta}(\gamma^{s})_{\delta\alpha}{\bf C}_{s}&\cr
& + {8\over 9\eta^{2}}(\bar{\lambda}\gamma^{ad}r)\xi_{d}^{\delta}(\gamma^{s})_{\delta\alpha}\partial_{s} - {8\over 3 \eta^{2}}(\bar{\lambda}\gamma^{ad}r)\xi_{d}^{\delta}(\gamma^{c})_{\delta\alpha}(\lambda\gamma_{cf}\lambda){\bf \Phi}^{f}& \cr
& & \identitythird
}
$$
where eqn. \drchata\ was used. One can already see from here that the first and third terms on the right-hand side of eqn. \identitythird\ will cancel the last BRST-exact term and the first term on the left-hand side of eqn. \firsteqnapcontinue, respectively. The last term of eqn. \identitythird\ can be put into the form
\eqnn \identityfourth
$$
\eqalignno{
- {8\over 3 \eta^{2}}(\bar{\lambda}\gamma^{ad}r)\xi_{d}^{\delta}(\gamma^{c})_{\delta\alpha}(\lambda\gamma_{cf}\lambda){\bf \Phi}^{f} &= {16\over 3\eta^{2}}(\bar{\lambda}\gamma^{ad}r)(\gamma^{dn}\lambda)_{\alpha}(\bar{\lambda}\gamma^{cn}\bar{\lambda})(\lambda\gamma_{cf}\lambda){\bf \Phi}^{f} & \cr
& - {8\over 3\eta^{2}}(\bar{\lambda}\gamma^{ac}r)(\gamma^{mn}\lambda)_{\alpha}(\bar{\lambda}\gamma_{mn}\bar{\lambda})(\lambda\gamma_{cf}\lambda){\bf \Phi}^{f} & \identityfourth
}
$$
which can be equivalently written as
\eqnn \identityfifth
$$
\eqalignno{
- {8\over 3 \eta^{2}}(\bar{\lambda}\gamma^{ad}r)\xi_{d}^{\delta}(\gamma^{c})_{\delta\alpha}(\lambda\gamma_{cf}\lambda){\bf \Phi}^{f} & = {8\over 3\eta}(\bar{\lambda}\gamma^{ad}r)(\gamma_{dn}\lambda)_{\alpha} {\bf \Phi}^{n} + {8\over 3\eta^{2}}(\bar{\lambda}\gamma^{ft}r)(\bar{\lambda}\gamma^{ac}\bar{\lambda})(\lambda\gamma_{cf}\lambda)(\gamma^{mn}\lambda)_{\alpha}(\bar{\lambda}\gamma_{mn}\bar{\lambda}){\bf \Sigma}_{t} &\cr
& + {4\over 3\eta^{2}}\phi(\gamma^{mn}\lambda)_{\alpha}(\bar{\lambda}\gamma_{mn}\bar{\lambda}){\bf \Phi}^{a} + {4\over 3\eta}(\bar{\lambda}\gamma^{at}r)(\gamma^{mn}\lambda)_{\alpha}(\bar{\lambda}\gamma_{mn}\bar{\lambda}){\bf \Sigma}_{t} &\cr
&= [Q, {4\over 3\eta }(\bar{\lambda}\gamma^{ad}\bar{\lambda})(\gamma_{dn}\lambda)_{\alpha}{\bf \Phi}^{n}] + {8\over 3 \eta^{2}}\phi (\bar{\lambda}\gamma^{ad}\bar{\lambda})(\gamma_{dn}\lambda)_{\alpha} {\bf \Phi}^{n} & \cr
& - {4\over 3\eta}(\bar{\lambda}\gamma^{ad}\bar{\lambda})(\gamma_{dn}\lambda)_{\alpha}(\lambda\gamma^{n}{\bf \Phi}) + {8\over 3\eta^{2}}(\lambda\gamma^{a}\xi_{f})(\bar{\lambda}\gamma^{ft}r)(\gamma^{mn}\lambda)_{\alpha}(\bar{\lambda}\gamma_{mn}\bar{\lambda}){\bf \Sigma}_{t} & \cr
& + {4\over 3\eta^{2}}\phi(\gamma^{mn}\lambda)_{\alpha}(\bar{\lambda}\gamma_{mn}\bar{\lambda}){\bf \Phi}^{a} & \identityfifth
}
$$
where we expressed ${\bf \Phi}^{a}$ as ${\bf \Phi}^{a} = (\bar{\lambda}\gamma^{ad}\bar{\lambda}){\bf \Sigma}_{d}$, and we used that $(\bar{\lambda}\gamma^{[ab}\bar{\lambda})(\bar{\lambda}\gamma^{cd]}r) = 0$. The use of the standard 11D identity $(\gamma^{ab})_{(\alpha\beta}(\gamma_{b})_{\epsilon\delta)} = 0$, then leads us to
\eqnn \identitysixth
$$
\eqalignno{
- {8\over 3 \eta^{2}}(\bar{\lambda}\gamma^{ad}r)\xi_{d}^{\delta}(\gamma^{c})_{\delta\alpha}(\lambda\gamma_{cf}\lambda){\bf \Phi}^{f} 
 &= [Q, -{2\over 3\eta}(\gamma_{dn}\lambda)_{\alpha}(\bar{\lambda}\gamma^{dn}\bar{\lambda}){\bf \Phi}^{a}]  + {8\over 3\eta^{2}}(\lambda\gamma^{a}\xi_{f})(\bar{\lambda}\gamma^{ft}r)(\gamma^{mn}\lambda)_{\alpha}(\bar{\lambda}\gamma_{mn}\bar{\lambda}){\bf \Sigma}_{t} &\cr
& - {1\over 3}(\gamma^{a}{\bf \Phi})_{\alpha} + {2\over 3\eta}(\lambda\gamma^{a}\xi_{n})(\gamma^{n}{\bf \Phi})+ {4\over 3\eta}(\lambda\gamma^{n})_{\alpha}(\lambda\gamma_{dn})^{\delta}{\bf \Phi}_{\delta}(\bar{\lambda}\gamma^{ad}\bar{\lambda})  & \cr
& & \identityfifth
}
$$
Likewise, the second terms on the right-hand side of eqns. \identitysecond, \identitythird\ combine into a single one term
\eqnn \identityseventh
$$
\eqalignno{
- {32\over 3\eta^{3}}(\bar{\lambda}\gamma^{ad}r)(\lambda\gamma_{dt}\lambda)(\bar{\lambda}\gamma^{pt}r)\xi_{p}^{\delta}(\gamma^{s})_{\delta\alpha}{\bf C}_{s} + {16\over 3\eta^{3}}\phi(\bar{\lambda}\gamma^{ad}r)\xi_{d}^{\delta}(\gamma^{s})_{\delta\alpha}{\bf C}_{s}&= -{32\over 3\eta^{2}}(\lambda\gamma^{a}[Q,\xi_{t}])(\bar{\lambda}\gamma^{pt}r)\xi_{p}^{\delta}(\gamma^{s})_{\delta\alpha}{\bf C}_{s} &\cr
& &\identityseventh
}
$$
Putting these results together, eqn. \firsteqnapcontinue\ simplifies to
\eqnn \identityeight
$$
\eqalignno{
{3\over 2}(\lambda\gamma^{a})_{\delta}\bigg[{\bf C}_{\alpha}\circ {\bf \Phi}^{\delta} + {\bf \Phi}^{\delta}\circ {\bf C}_{\alpha} - &{2\over 3\eta}\xi_{n}^{\delta}(\gamma^{n}{\bf \Phi}) - {8\over 3\eta^{2}}\xi_{f}^{\delta}(\bar{\lambda}\gamma^{ft}r)(\gamma^{mn}\lambda)_{\alpha}(\bar{\lambda}\gamma_{mn}\bar{\lambda}){\bf \Sigma}_{t} & \cr
& + {32\over 3\eta^{3}}[Q,\xi_{t}^{\delta}](\bar{\lambda}\gamma^{pt}r)\xi_{p}^{\delta}(\gamma^{s})_{\delta\alpha}{\bf C}_{s}\bigg] = (\lambda\gamma^{a})_{\delta}{\bf h}_{\alpha}{}^{\delta} & \identityseventh
}
$$
Thus, one straightforwardly concludes that
\eqnn \solufive
$$
\eqalignno{
{\bf h}_{\alpha}{}^{\delta} &= {3\over 2}\bigg[{\bf C}_{\alpha}\circ {\bf \Phi}^{\delta} + {\bf \Phi}^{\delta}\circ {\bf C}_{\alpha} - {2\over 3\eta}\xi_{n}^{\delta}(\gamma^{n}{\bf \Phi}) - {8\over 3\eta^{2}}\xi_{f}^{\delta}(\bar{\lambda}\gamma^{ft}r)(\gamma^{mn}\lambda)_{\alpha}(\bar{\lambda}\gamma_{mn}\bar{\lambda}){\bf \Sigma}_{t} & \cr
& + {32\over 3\eta^{3}}[Q,\xi_{t}^{\delta}](\bar{\lambda}\gamma^{pt}r)\xi_{p}^{\delta}(\gamma^{s})_{\delta\alpha}{\bf C}_{s}\bigg] & \solufive
}
$$

\appendix{C}{Cubic Physical Operators}

\seclab\appendixb

A procedure analogous to that developed in Appendix B for quadratic propagators can be applied to cubic operators. One starts with the definition $\tilde{{\bf C}}_{\alpha\beta\delta} = {9\over 2}{\bf C}_{(\alpha} \circ {\bf C}_{\beta} \circ {\bf C}_{\delta)}$, and plugs it into the right-hand side of the relation \cubicopone 
\eqnn \calphabetadeltareplace
$$
\eqalignno{
 {9\over 2}[Q , {\bf C}_{(\alpha} \circ {\bf C}_{\beta} \circ {\bf C}_{\delta)}] &= 3{\bf C}_{(\alpha}\circ[Q, {\bf C}_{\beta\delta)}] + 3[Q, {\bf C}_{(\alpha}] \circ {\bf C}_{\beta\delta)} & \cr
 &= -3D_{(\alpha}{\bf C}_{\beta}\circ {\bf C}_{\delta)} - 6{\bf C}_{(\alpha}(\lambda\gamma^{a})_{\beta}{\bf C}_{a\delta)} - {3\over 2}(\gamma^{a})_{(\beta\delta}{\bf C}_{\alpha)}\circ {\bf C}_{a} & \cr
 & - D_{(\alpha}{\bf C}_{\beta\delta)} - 3(\gamma^{a}\lambda)_{(\alpha}{\bf C}_{a}\circ {\bf C}_{\beta\delta)} & \cr
  &= -3D_{(\alpha}{\bf C}_{\beta\delta)} - 2(\gamma^{a})_{(\alpha\beta}{\bf C}_{a\delta)} - {3\over 2}(\gamma^{a})_{(\beta\delta}{\bf C}_{\alpha)}\circ {\bf C}_{a}& \cr
  & - {2 \over \eta}(\lambda\gamma^{pqr}\gamma^{a})_{(\beta}(\bar{\lambda}\gamma_{pq}\bar{\lambda})(\lambda\gamma_{r})_{\alpha}{\bf C}_{a\delta)} &\cr
 & -6(\lambda\gamma^{a})_{(\alpha}{\bf C}_{\beta}\circ{\bf C}_{a\delta)}  - 3(\gamma^{a}\lambda)_{(\alpha}{\bf C}_{a}\circ {\bf C}_{\beta\delta)}& \calphabetadeltareplace
}
$$
Recalling the definition of ${\bf C}_{a\alpha}$
\eqnn \calphabetapartivapp
$$
\eqalignno{
{\bf C}_{a\alpha} &=  {3\over 2}{\bf C}_{\alpha}\circ {\bf C}_{a} + {3\over 4}[{\bf C}_{a}, {\bf C}_{\alpha}] + {1\over 4\eta}(\lambda\gamma^{apq}\gamma^{b})_{\alpha}(\bar{\lambda}\gamma^{pq}\bar{\lambda}){\bf C}_{b} & \calphabetapartivapp
}
$$
The substitution of eqn. \commchatalphachatapartii\ into eqn. \calphabetapartivapp, results in the following alternative expression for ${\bf C}_{a\alpha}$
\eqnn \theorem
$$
\eqalignno{
{\bf C}_{a\alpha} &=  {3\over 2}{\bf C}_{\alpha}\circ {\bf C}_{a} + {3\over 2\eta}\xi_{a}^{\beta}\left[{\bf C}_{\alpha},[Q, {\bf C}_{\beta}]\right] & \theorem 
}
$$
Therefore,
\eqnn \eqforca
$$
\eqalignno{
[Q , \tilde{{\bf C}}_{\alpha\beta\delta}] &= -3D_{(\alpha}{\bf C}_{\beta\delta)} - 3(\gamma^{a})_{(\alpha\beta}{\bf C}_{a\delta)} + {3\over 2\eta}(\gamma^{a})_{(\alpha\beta}\xi_{a}^{\epsilon}[{\bf C}_{\delta)}, [Q, {\bf C}_{\epsilon}]]& \cr
  & -3(\lambda\gamma^{a})_{(\alpha} \bigg[{1 \over \eta}(\lambda\gamma^{pqa}\gamma^{r})_{\beta}(\bar{\lambda}\gamma_{pq}\bar{\lambda}){\bf C}_{r\delta)}  +  2{\bf C}_{\beta}\circ{\bf C}_{a\delta)}  + {\bf C}_{a}\circ {\bf C}_{\beta\delta)}\bigg] &\eqforca
}
$$
Using the Jacobi identity and the fact that $[Q, {\bf C}_{\alpha}] = (\lambda\gamma^{a})_{\alpha}F_{a}$, where
\eqnn \fdefinition
$$
\eqalignno{
F_{a} &= {1\over 3\eta}(\bar{\lambda}\gamma^{bc}d)(\lambda\gamma_{abc}\lambda) - {\bf C}_{a} & \fdefinition
}
$$
one arrives at
\eqnn \eqforcaimp
$$
\eqalignno{
[Q , \tilde{{\bf C}}_{\alpha\beta\delta}] &= -3D_{(\alpha}{\bf C}_{\beta\delta)} - 3(\gamma^{a})_{(\alpha\beta}{\bf C}_{a\delta)} - {3\over 2\eta}(\lambda\gamma^{b})_{(\alpha}(\gamma^{a})_{\beta\delta)}[\xi_{a}^{\epsilon}, F_{b}]{\bf C}_{\epsilon} + {3\over 2\eta}(\gamma^{a})_{(\alpha\beta}\xi_{a}^{\epsilon}[Q,[{\bf C}_{\delta)},{\bf C}_{\epsilon}]] & \cr
& -3(\lambda\gamma^{a})_{(\alpha} \bigg[{1 \over \eta}(\lambda\gamma^{pqa}\gamma^{r})_{\beta}(\bar{\lambda}\gamma_{pq}\bar{\lambda}){\bf C}_{r\delta)}  +  2{\bf C}_{\beta}\circ{\bf C}_{a\delta)} + {\bf C}_{a}\circ {\bf C}_{\beta\delta)}\bigg] &\cr
&= -3D_{(\alpha}{\bf C}_{\beta\delta)} - 3(\gamma^{a})_{(\alpha\beta}{\bf C}_{a\delta)} - {3\over 2\eta}(\lambda\gamma^{b})_{(\alpha}(\gamma^{a})_{\beta\delta)}[\xi_{a}^{\epsilon}, F_{b}]{\bf C}_{\epsilon} &\cr
&+ {1\over 2\eta}(\gamma^{a})_{(\alpha\beta}\xi_{a}^{\epsilon}(\lambda\gamma^{r})_{\delta)}[Q, {1\over \eta}(\bar{\lambda}\gamma^{pq}\bar{\lambda})(\gamma_{pqr})_{\epsilon}{}^{\kappa}{\bf C}_{\kappa}] &\cr
& - {1\over 2\eta}(\gamma^{a})_{(\alpha\beta}(\gamma_{pqr})_{\delta)}{}^{\kappa}\xi_{a}^{\epsilon}(\lambda\gamma^{r})_{\epsilon}[Q, {1\over \eta}(\bar{\lambda}\gamma^{pq}\bar{\lambda}){\bf C}_{\kappa}] & \cr
& + {2\over 3\eta}(\gamma^{a})_{(\alpha\beta}(\lambda\gamma_{f})_{\delta)}\xi^{\epsilon}_{a}[Q,{1\over \eta^{2}}(\lambda\gamma^{cd}w)(\bar{\lambda}\gamma_{cd}\bar{\lambda})(\bar{\lambda}\gamma^{fg}\bar{\lambda})(\lambda\gamma_{g})_{\epsilon} ] &\cr
& -3(\lambda\gamma^{a})_{(\alpha} \bigg[{1 \over \eta}(\lambda\gamma^{pqa}\gamma^{r})_{\beta}(\bar{\lambda}\gamma_{pq}\bar{\lambda}){\bf C}_{r\delta)}  +  2{\bf C}_{\beta}\circ{\bf C}_{a\delta)} + {\bf C}_{a}\circ {\bf C}_{\beta\delta)}\bigg] & \eqforcaimp
}
$$
where we used eqn. \commutatorcalphacbetafinal. The fifth term on the right-hand side of eqn. \eqforcaimp\ can be written in the equivalent form
\eqnn \fifthtermequivalent
$$
\eqalignno{
- {1\over 2\eta}(\gamma^{a})_{(\alpha\beta}(\gamma_{pqr})_{\delta)}{}^{\kappa}\xi_{a}^{\epsilon}(\lambda\gamma^{r})_{\epsilon}[Q, {1\over \eta}(\bar{\lambda}\gamma^{pq}\bar{\lambda}){\bf C}_{\kappa}] &= -{1\over 4}(\gamma^{a})_{(\alpha\beta}(\gamma_{pqa})_{\delta)}{}^{\kappa}[Q, {1\over \eta}(\bar{\lambda}\gamma^{pq}\bar{\lambda}){\bf C}_{\kappa}] & \cr
& + {1\over 2\eta}(\gamma^{a})_{(\alpha\beta}(\lambda\gamma_{at}\lambda)(\bar{\lambda}\gamma^{rt}\bar{\lambda})(\gamma_{pqr})_{\delta)}{}^{\kappa}[Q, {1\over \eta}(\bar{\lambda}\gamma^{pq}\bar{\lambda}){\bf C}_{\kappa}]  &\cr
& & \fifthtermequivalent
}
$$
which, after using the 11D identity $(\gamma^{ab})_{(\alpha\beta}(\gamma_{b})_{\delta\epsilon} = 0$, reduces to
\eqnn \fifthtermequivalenttwo
$$
\eqalignno{
- {1\over 2\eta}(\gamma^{a})_{(\alpha\beta}(\gamma_{pqr})_{\delta)}{}^{\kappa}\xi_{a}^{\epsilon}(\lambda\gamma^{r})_{\epsilon}[Q, {1\over \eta}(\bar{\lambda}\gamma^{pq}\bar{\lambda}){\bf C}_{\kappa}] &= 
[Q, -{1\over 4\eta}(\gamma^{a})_{(\alpha\beta}(\gamma_{pqa})_{\delta)}{}^{\kappa}(\bar{\lambda}\gamma^{pq}\bar{\lambda}){\bf C}_{\kappa}] &\cr
& -{2\over \eta}(\lambda\gamma^{a})_{(\alpha}(\gamma_{at}\lambda)_{\beta)}(\bar{\lambda}\gamma^{rt}\bar{\lambda})(\gamma_{pqr})_{\delta)}{}^{\kappa}[Q, {1\over \eta}(\bar{\lambda}\gamma^{pq}\bar{\lambda}){\bf C}_{\kappa}] &
}
$$
Therefore, one learns that
\eqnn \cubicoperatorequation
$$
\eqalignno{
[Q , \tilde{{\bf C}}_{\alpha\beta\delta}] 
&= -3D_{(\alpha}{\bf C}_{\beta\delta)} - 3(\gamma^{a})_{(\alpha\beta}{\bf C}_{a\delta)} &\cr
& + [Q, -{1\over 4\eta}(\gamma^{a})_{(\alpha\beta}(\gamma_{pqa})_{\delta)}{}^{\kappa}(\bar{\lambda}\gamma^{pq}\bar{\lambda}){\bf C}_{\kappa}] &\cr
& -3(\lambda\gamma^{a})_{(\alpha} \bigg[2{\bf C}_{\beta}\circ{\bf C}_{a\delta)} + {\bf C}_{a}\circ {\bf C}_{\beta\delta)} + {1 \over \eta}(\lambda\gamma^{pqa}\gamma^{r})_{\beta}(\bar{\lambda}\gamma_{pq}\bar{\lambda}){\bf C}_{r\delta)} &\cr
& + {1\over 2\eta}(\gamma^{b})_{\beta\delta)}[\xi_{b}^{\epsilon}, F_{a}]{\bf C}_{\epsilon} - {1\over 6\eta}(\gamma^{f})_{(\alpha\beta}\xi_{f}^{\epsilon}[Q, {1\over \eta}(\bar{\lambda}\gamma^{pq}\bar{\lambda})(\gamma_{pqa})_{\epsilon}{}^{\kappa}{\bf C}_{\kappa}]&\cr
& + {2\over 3\eta}(\gamma_{at}\lambda)_{\beta)}(\bar{\lambda}\gamma^{rt}\bar{\lambda})(\gamma_{pqr})_{\delta)}{}^{\kappa}[Q, {1\over \eta}(\bar{\lambda}\gamma^{pq}\bar{\lambda}){\bf C}_{\kappa}] &\cr
& - {2\over 9\eta}(\gamma^{f})_{(\beta\delta}\xi^{\epsilon}_{f}[Q,{1\over \eta^{2}}(\lambda\gamma^{cd}w)(\bar{\lambda}\gamma_{cd}\bar{\lambda})(\bar{\lambda}\gamma_{ag}\bar{\lambda})(\lambda\gamma^{g})_{\epsilon}] \bigg] & \cubicoperatorequation
}
$$
Eqn. \cubicoperatorequation\ becomes eqn. \cubicopone\ if one makes the following identifications
\eqnn \idenone
\eqnn \identwo
$$
\eqalignno{
{\bf C}_{\alpha\beta\delta} &= \tilde{{\bf C}}_{\alpha\beta\delta} + {1\over 4\eta}(\gamma^{a})_{(\alpha\beta}(\gamma_{pqa})_{\delta)}{}^{\kappa}(\bar{\lambda}\gamma^{pq}\bar{\lambda}){\bf C}_{\kappa} & \idenone \cr
{\bf C}_{a\beta\delta} &= 2{\bf C}_{\beta}\circ{\bf C}_{a\delta)} + {\bf C}_{a}\circ {\bf C}_{\beta\delta)} + {1 \over \eta}(\lambda\gamma^{pqa}\gamma^{r})_{\beta}(\bar{\lambda}\gamma_{pq}\bar{\lambda}){\bf C}_{r\delta)} &\cr
& + {1\over 2\eta}(\gamma^{b})_{\beta\delta)}[\xi_{b}^{\epsilon}, F_{a}]{\bf C}_{\epsilon} - {1\over 6\eta}(\gamma^{f})_{(\alpha\beta}\xi_{f}^{\epsilon}[Q, {1\over \eta}(\bar{\lambda}\gamma^{pq}\bar{\lambda})(\gamma_{pqa})_{\epsilon}{}^{\kappa}{\bf C}_{\kappa}]&\cr
& + {2\over 3\eta}(\gamma_{at}\lambda)_{\beta)}(\bar{\lambda}\gamma^{rt}\bar{\lambda})(\gamma_{pqr})_{\delta)}{}^{\kappa}[Q, {1\over \eta}(\bar{\lambda}\gamma^{pq}\bar{\lambda}){\bf C}_{\kappa}] &\cr
& - {2\over 9\eta}(\gamma^{f})_{(\beta\delta}\xi^{\epsilon}_{f}[Q,{1\over \eta^{2}}(\lambda\gamma^{cd}w)(\bar{\lambda}\gamma_{cd}\bar{\lambda})(\bar{\lambda}\gamma_{ag}\bar{\lambda})(\lambda\gamma^{g})_{\epsilon}] & \identwo
}
$$


\appendix{D}{Non-Linear Physical Operators in Action}

\seclab\appendixb
In this Appendix, we provide a detailed derivation of eqns. \chatalphabonpsi\ and \chatabonpsi. Let us start with ${\bf C}_{a \alpha}$, and reprint its definition here
\eqnn \defchataalphareprint
$$
\eqalignno{
{\bf C}_{a\alpha} &= {3\over 4}({\bf C}_{a}\circ {\bf C}_{\alpha} + {\bf C}_{\alpha} \circ {\bf C}_{a}) + {1\over 4\eta}(\lambda\gamma^{apq}\gamma^{b})_{\alpha}(\bar{\lambda}\gamma_{pq}\bar{\lambda}){\bf C}_{b} & \defchataalphareprint
}
$$
To compute the action of this operator on $U^{(3)}$, one needs to make use of the relations \chatalphaonpsi\ and \chataonpsi. In doing so, the first term of eqn. \defchataalphareprint\ can be written as 
\eqnn \firsttermcaalphaonpsi
$$
\eqalignno{
{\bf C}_{\alpha}({\bf C}_{a}(U^{(3)})) &= {\bf C}_{\alpha}(C_{a} + (\lambda\gamma_{ac}\lambda)s^{c} - Q\rho_{a}) & \cr
 &= {2\over 3}C_{a\alpha} + (\lambda\gamma^{r})_{\alpha}\bigg[{2\over 3\eta}(\lambda\gamma_{pqr}C_{a})(\bar{\lambda}\gamma^{pq}\bar{\lambda})\bigg] + {2\over 3}(\gamma_{ac}\lambda)_{\alpha}s^{c} + (\lambda\gamma_{ac}\lambda){\bf C}_{\alpha}s^{c}& \cr
 & - [{\bf C}_{\alpha}, Q]\rho_{a} - Q({\bf C}_{\alpha}\rho_{a}) & \cr
&= {2\over 3}C_{a\alpha} + (\lambda\gamma^{r})_{\alpha}\bigg[{2\over 3\eta}(\lambda\gamma_{pqr}C_{a})(\bar{\lambda}\gamma^{pq}\bar{\lambda}) - {\bf C}_{r}\rho_{a}\bigg] + {2\over 3}(\gamma_{ac}\lambda)_{\alpha}s^{c} + (\lambda\gamma_{ac}\lambda){\bf C}_{\alpha}s^{c} &\cr
& -{1\over 3}D_{\alpha}\rho_{a} - Q({\bf C}_{\alpha}\rho_{a}) &\firsttermcaalphaonpsi
}
$$

The second term in \defchataalphareprint, in turn, can be cast as
\eqnn \secondtermcaalphaonpsi
$$
\eqalignno{
{\bf C}_{a}({\bf C}_{\alpha}(U^{(3)})) &= {\bf C}_{a}(C_{\alpha} + (\lambda\gamma^{b})_{\alpha}\rho_{b}) & \cr
&= -{2\over \eta}\xi^{\delta}_{a}({D_{\delta}\over 3} + Q{\bf C}_{\delta})\left(C_{\alpha} + (\lambda\gamma^{b})_{\alpha}\rho_{b}\right) & \secondtermcaalphaonpsi
}
$$
The use of eqn. \chatalphabetaonpsi\ then enables one to write
\eqnn \chatachatalphaonpi
$$
\eqalignno{
{\bf C}_{a}({\bf C}_{\alpha}(U^{(3)})) &= -{2\over \eta}\xi_{a}^{\delta}\bigg[{1\over 3}D_{\delta}C_{\alpha} + {1\over 3}(\lambda\gamma^{b})_{\alpha}D_{\delta}\rho_{b} + {2\over 3}QC_{\alpha\delta} + {1\over 3}(\gamma^{b})_{\alpha\delta}Q\rho_{b} + (\lambda\gamma^{b})_{\alpha}Q{\bf C}_{\delta}\rho_{b} & \cr
& + (\lambda\gamma_{r})_{\delta}Q\bigg({2\over 3\eta}(\lambda\gamma^{pqr}C_{\alpha})(\bar{\lambda}\gamma_{pq}\bar{\lambda}) + {1\over 3\eta}(\lambda\gamma^{pqr}\gamma^{b})_{\alpha}(\bar{\lambda}\gamma_{pq}\bar{\lambda})\rho_{b}\bigg)\bigg] &\cr
&= {4\over3\eta}(\lambda\gamma^{b})_{\alpha}\xi_{a}^{\delta}C_{b\delta} + {2\over 3}C_{a\alpha} - {4\over 3\eta}(\lambda\gamma_{ak}\lambda)(\bar{\lambda}\gamma^{bk}\bar{\lambda})C_{b\alpha} + {2\over 3\eta}\xi_{a}^{\delta}(\gamma^{b})_{\alpha\delta}C_{b} + {2\over 3\eta}D_{\alpha}\xi_{a}^{\delta}C_{\delta} &\cr
& -{2\over 3\eta}(\lambda\gamma^{b})_{\alpha}\xi_{a}^{\delta}D_{\delta}\rho_{b} - {2\over 3\eta}(\gamma^{b})_{\alpha\delta}\xi_{a}^{\delta}Q\rho_{b} + (\lambda\gamma^{b})_{\alpha}(-{2\over \eta}\xi_{a}^{\delta}Q{\bf C}_{\delta}\rho_{b}) & \cr
& - Q\bigg({2\over 3\eta}(\lambda\gamma_{pqa}C_{\alpha})(\bar{\lambda}\gamma^{pq}\bar{\lambda}) + {1\over 3\eta}(\lambda\gamma_{pqa}\gamma^{b})_{\alpha}(\bar{\lambda}\gamma^{pq}\bar{\lambda})\rho_{b}\bigg) &\cr
& + {2\over \eta}(\lambda\gamma_{ak}\lambda)(\bar{\lambda}\gamma^{rk}\bar{\lambda})Q\bigg({2\over 3\eta}(\lambda\gamma_{pqr}C_{\alpha})(\bar{\lambda}\gamma^{pq}\bar{\lambda}) + {1\over 3\eta}(\lambda\gamma_{pqr}\gamma^{b})_{\alpha}(\bar{\lambda}\gamma^{pq}\bar{\lambda})\rho_{b}\bigg) & \chatachatalphaonpi
}
$$

The third term of eqn. \defchataalphareprint\ can easily be computed to be
\eqnn \thirdtermcacalphaonpsi
$$
\eqalignno{
{1\over 4\eta}(\lambda\gamma^{apq}\gamma^{b})_{\alpha}(\bar{\lambda}\gamma_{pq}\bar{\lambda}){\bf C}_{b}(U^{(3)}) 
&= -{1\over 2\eta}\xi_{a}^{\delta}(\gamma^{b})_{\delta\alpha}C_{b} + {1\over 2\eta}\xi^{\delta}_{a}(\gamma^{b})_{\delta\alpha}Q\rho_{b} + {1\over \eta}(\gamma_{aq}\lambda)_{\alpha}(\lambda\gamma_{pr}\lambda)(\bar{\lambda}\gamma^{pq}\bar{\lambda})s^{r} &\cr
& - {1\over 2\eta}(\lambda\gamma_{ar}\lambda)(\gamma^{pq}\lambda)_{\alpha}(\bar{\lambda}\gamma_{pq}\bar{\lambda})s^{r} &\thirdtermcacalphaonpsi
}
$$
Putting eqns. \firsttermcaalphaonpsi, \chatachatalphaonpi, \thirdtermcacalphaonpsi\ together as dictated by eqn. \defchataalphareprint, one reproduces eqn. \chatalphabonpsi.

\medskip
In a similar fashion, one can compute the action of ${\bf C}_{ab} = -{3\over 2}{\bf C}_{[a}\circ {\bf C}_{b]}$ on $U^{(3)}$:
\eqnn \chatachatbonpsi
$$
\eqalignno{
{\bf C}_{a}({\bf C}_{b}U^{(3)}) &= {1\over \eta}(\lambda\gamma^{agh})^{\alpha}(\bar{\lambda}\gamma_{gh}\bar{\lambda})\bigg[{1\over 3}D_{\alpha}C_{b}  + {2\over 3}QC_{b\alpha} + {1\over 3}(\lambda\gamma_{bm}\lambda)D_{\alpha}s^{m} + (\lambda\gamma_{bm}\lambda)Q({\bf C}_{\alpha}s^{m}) &\cr
& + {2\over 3}(\lambda\gamma^{r})_{\alpha}Q\tilde{f}_{rb} + {2\over 3}(\gamma_{bm}\lambda)_{\alpha}Qs^{m} + {1\over 3}(\lambda\gamma^{t})_{\alpha}\partial_{t}\rho_{b}\bigg] & \chatachatbonpsi
}
$$
where we used eqn. \chataonpsi, and that $\tilde{f}_{rb} = {1\over \eta}(\lambda\gamma_{rpq}C_{b})(\bar{\lambda}\gamma^{pq}\bar{\lambda}) - {3\over 2}{\bf C}_{r}\rho_{b}$. The use of the equation of motion \eomeight\ then yields 
\eqnn \chatachatbonpsieom
$$
\eqalignno{
{\bf C}_{a}({\bf C}_{b}(U^{(3)}) &= {1\over 3}\partial_{b}\rho_{a} - {1\over 3}\partial_{a}\rho_{b} - {2\over 3}Qf_{ab} - {2\over 3}C_{ab} + Q{\bf C}_{a}\rho_{b}) + (\lambda\gamma^{ah}\lambda)\bigg[{2\over \eta}(\bar{\lambda}\gamma^{sh}\bar{\lambda})\bigg({2\over 3}C_{sb} + {2\over 3}Qf_{sb} &\cr
& + {1\over 3}\partial_{s}\rho_{b} - Q({\bf C}_{s}\rho_{b})\bigg)\bigg] + (\lambda\gamma^{bt}\lambda)\bigg[-{1\over 3\eta}(\lambda\gamma^{agh})^{\alpha}h_{\alpha}{}^{t}(\bar{\lambda}\gamma_{gh}\bar{\lambda}) + {1\over 3\eta}(\lambda\gamma^{agh})^{\alpha}D_{\alpha}s^{t}(\bar{\lambda}\gamma^{gh}\bar{\lambda}) &\cr
& + {1\over \eta}(\lambda\gamma^{agh})^{\alpha}(\bar{\lambda}\gamma_{gh}\bar{\lambda})Q(\hat{C}_{\alpha}s^{t})\bigg] + (\lambda\gamma^{ab})^{\alpha}\bigg[{1\over \eta}(\gamma^{fgh}\lambda)_{\alpha}(\bar{\lambda}\gamma_{gh}\bar{\lambda})\bigg(-{2\over 3}\phi^{f} + {2\over 3}Qs^{f}\bigg)\bigg] &\cr
& & \chatachatbonpsi
}
$$
After multiplying by a factor of $-{3\over 2}$ on both sides, and antisymmetrizing in $[a,b]$, it is easy to see that eqn. \chatachatbonpsi\ coincides with eqn. \chatabonpsi.
\listrefs

\bye